\documentclass[12pt,preprint]{aastex}

\newcommand\arcpt{${{\lower3pt\hbox{$^{\prime\prime}$}}\atop{\raise4pt\hbox{.}}}$}

\newcommand\mjup{$M_{Jup}$}
\newcommand\sdeg{$^\circ$}

\slugcomment{to appear in the {\it Astrophysical Journal}}

\shorttitle{Two Suns in The Sky}
\shortauthors{Raghavan et al.}

\begin{document}

\title{Two Suns in The Sky: \\ Stellar Multiplicity in Exoplanet
Systems}

\author{Deepak Raghavan, Todd J. Henry}

\affil{Georgia State University, Atlanta, GA 30302-4106}

\author{Brian D. Mason}

\affil{US Naval Observatory, 3450 Massachusetts Avenue NW, 
Washington DC 20392-5420}

\author{John P. Subasavage, Wei-Chun Jao, Thom D. Beaulieu}

\affil{Georgia State University, Atlanta, GA 30302-4106}

\author{Nigel C. Hambly}

\affil{Institute for Astronomy, School of Physics, University of Edinburgh, 
Royal Observatory, Blackford Hill, Edinburgh EH9 3HJ, Scotland, UK}

\email{raghavan@chara.gsu.edu}

\begin{abstract}

We present results of a reconnaissance for stellar companions to all
131 radial-velocity-detected candidate extrasolar planetary systems
known as of July 1, 2005.  Common proper motion companions were
investigated using the multi-epoch STScI Digitized Sky Surveys, and
confirmed by matching the trigonometric parallax distances of the
primaries to companion distances estimated photometrically.  We also
attempt to confirm or refute companions listed in the Washington
Double Star Catalog, the Catalogs of Nearby Stars Series by Gliese
and Jahrei{\ss}, in Hipparcos results, and in \citet{Duq1991}.

Our findings indicate that a lower limit of 30 (23\%) of the 131
exoplanet systems have stellar companions.  We report new stellar
companions to HD 38529 and HD 188015, and a new candidate companion to
HD 169830.  We confirm many previously reported stellar companions,
including six stars in five systems, that are recognized for the first
time as companions to exoplanet hosts.  We have found evidence that 20
entries in the Washington Double Star Catalog are not gravitationally
bound companions.  At least three (HD 178911, 16 Cyg B, and HD
219449), and possibly five (including HD 41004 and HD 38529), of the
exoplanet systems reside in triple star systems.  Three exoplanet
systems (GJ 86, HD 41004, and $\gamma$ Cep) have potentially close-in
stellar companions, with planets at $\sim$ Mercury to Mars distances
from the host star and stellar companions at projected separations of
$\sim$ 20 AU, similar to the Sun--Uranus distance.  Finally, two of
the exoplanet systems contain white dwarf companions.  This
comprehensive assessment of exoplanet systems indicates that solar
systems are found in a variety of stellar multiplicity environments --
singles, binaries, and triples; and that planets survive the
post-main-sequence evolution of companion stars.

\end{abstract}


\section{Introduction}

The hunt for planets outside our solar system has revealed 161
candidate planets in 137 stellar systems as of July 1, 2005, with 18
of these systems containing multiple planets.  After the initial
flurry of ``Hot Jupiter'' discoveries --- primarily a selection effect
due to two factors: (1) the nascent effort was biased toward discovery
of short period systems, and (2) massive planets induce more readily
detected radial velocity variations --- it is now believed that the
more massive planets preferentially lie farther away from the primary
\citep{Udry2004,Mar2005b}, perhaps leaving the space closer to the
star for the harder to detect terrestrial planets.  Through these
discoveries, we are now poised to gain a better understanding of the
environments of exoplanet systems and compare them to our Solar
System.

Our effort in this paper is focused on a key parameter of planetary
systems --- the stellar multiplicity status of exoplanet hosts.  We
address questions such as: (1) Do planets preferentially occur in
single star systems (like ours), or do they commonly occur in multiple
star systems as well?  (2) For planets residing in multiple star
systems, how are the planetary orbits related to stellar separations?  (3)
What observational limits can we place on disk or orbit disruptions in
multi-star planetary systems?  This study contributes to the broader
subjects of planetary system formation, evolution and stability
through a better understanding of the environments of exoplanet
systems.

Stellar multiplicity among exoplanet systems was first studied by
\citet{Pat2002}, who looked at the first 11 exoplanet systems
discovered and reported two binaries and one triple system.
\citet{Luh2002} conducted an adaptive optics survey looking for
stellar and sub-stellar companions to 25 exoplanet hosts and reported
null results. More recently, \citet{Egg2004} and \citet{Udry2004}
reported 15 exoplanet systems with stellar companions in a
comprehensive assessment, and additional companions have been reported
for several specific systems \citep{Mug2004a,Mug2004b,Mug2005b}.  Our
effort confirms many of these previously reported systems, reports two
new companions, identifies an additional candidate, and recognizes,
for the first time, one triple and four binary exoplanet systems
(these are known stellar companions, but previously not noted to
reside in exoplanet systems).

\section {Sample \& Companion Search Methodology}
\label{sec:Sample}

Our sample includes all known exoplanet systems detected by radial
velocity techniques as of July 1, 2005.  We primarily used the
Extrasolar Planets Catalog, maintained by Jean Schneider at the Paris
Observatory\footnote{$http://vo.obspm.fr/exoplanetes/encyclo/catalog.php$},
to build our sample list for analyses.  To ensure completeness, we
cross-checked this list with the California \& Carnegie Planet Search
Catalog\footnote{$http://exoplanets.org/$}.  Our sample excludes
planets discovered via transits and gravitational lensing, as these
systems are very distant, with poor or no parallax and magnitude
information for the primaries.  In addition, these systems can not be
observed for stellar companions in any meaningful way.  We also
exclude a radial velocity detected system, HD 219542, identified by
\citet{Egg2004} as an exoplanet system with multiple stars, but since
confirmed as a false planet detection by its discoverers
\citep{Des2004}. The final sample comprises 155 planets in 131
systems. This list is included in Table~\ref{Sample} along with
companion detection information, as described below.

Several efforts were carried out to gather information on stellar
companions to exoplanet stars.  To identify known or claimed
companions, we checked available sources listing stellar companions
--- the Washington Double Star Catalog (WDS), the Hipparcos Catalog,
the Catalog of Nearby Stars (\citealt{CNS1969,CNS1979,CNS1991},
hereafter CNS) and \citet{Duq1991}.  We also visually inspected the
STScI Digitized Sky Survey (DSS) multi-epoch frames for the sky around
each exoplanet system to investigate reported companions and to
identify new common proper motion (CPM) companion candidates.  We then
confirmed or refuted many candidates through photometric distance
estimates using plate magnitudes from SuperCOSMOS, optical CCD
magnitudes from the CTIO 0.9m and 1.0m telescopes, and infrared
magnitudes from 2MASS.  The origin and status of each companion is
summarized in Table~\ref{results} and described in
\S\ref{sec:NotESys}.

Table~\ref{Sample} lists each target star in our sample, sequenced
alphabetically by name, and identifies all known and new companions.
The first column is the exoplanet host star's name (HD when available,
otherwise BD or GJ name).  The second and third columns give the
proper motion magnitude (in seconds of arc per year) and direction (in
degrees) of the star, mostly from Hipparcos.  The fourth and
fifth columns specify the observational epochs of the DSS images
blinked to identify CPM companion candidates.  The sixth column lists
the total proper motion, in seconds of arc, of the exoplanet host
during the time interval between the two observational epochs of the
DSS plates.  The seventh column identifies whether the proper motion
of the star was detectable in the DSS frames, allowing the
identification of CPM candidates.  The entries ``YES'' and ``NO'' are
self-explanatory, and ``MAR'' identifies that the proper motion was
marginally detectable.  Systems with very little proper motion or a
brief separation between plate epochs could not be searched
effectively (see \S\ref{sec:DSS}).  The eighth column specifies
companions identified via CPM, and the ninth column specifies
companions listed in the sources mentioned above or in other refereed
papers.  A ``?''  following the companion ID indicates that the source
remains a candidate, and could not be confirmed or refuted with
confidence.  The absence of a question mark indicates that the
companion is confirmed.

Each reference we used for the companion search is described in the 
subsections below.

\subsection {STScI Digitized Sky Survey (DSS)}
\label{sec:DSS}

We downloaded multi-epoch images of the sky around each exoplanet
primary from the STScI Digitized Sky
Survey\footnote{$http://stdatu.stsci.edu/cgi-bin/dss\_form$}. The
images of these surveys are based on photographic data obtained using
the Oschin Schmidt Telescope on Palomar Mountain and the UK Schmidt
Telescope in Australia. We typically extracted 10$\arcmin$ square
images at two epochs centered on an exoplanet host star.  The range of
time interval between the epochs for a given target is 3.1 years to
46.2 years.  Figure~\ref{DSS-hist} shows a histogram of the number of
systems per time interval bin for our sample.

We identified CPM companion candidates by eye, by blinking the two
epoch frames.  In general, primaries with a total proper motion of
$\ge$ 3$\arcsec$ were effectively searched, while those with a total
proper motion in the range of 2 $-$ 3 $\arcsec$ were marginally
searched, and stars with $\le$ 2 $\arcsec$ total proper motion could
not be searched for companions using this method.  Exceptions to these
ranges exist, and are due to poorly matched astrometric fields caused
by specific issues with the plate images, such as saturation around
the primary, distribution of background stars in the frames,
brightness of the companion and its proximity to the primary, and the
relative rotation between the frames.  The 3$\arcsec$ detection limit
corresponds to a proper motion range of 0\farcs1 yr$^{-1}$ to 1\farcs0
yr$^{-1}$ with a median value of 0\farcs2 yr$^{-1}$ for the time
intervals sampled.  Additionally, this method favors the detection of
wide companions because bright primaries saturate the surrounding
region out to many seconds of arc, and prevent companion detection
within a $\sim$ 15$\arcsec$ $-$ 30$\arcsec$ radius, depending on
source brightness.  At the median distance of 35.6 pc for our sample,
this translates to a minimum projected distance of $\sim$ 500 $-$ 1000
AU.  However, some bright companions can be picked up much closer, due
to twin diffraction spikes or an anomalous PSF compared to other stars
in the field.  For an outer limit, the 10$\arcmin$ image gives us a
radius of 5$\arcmin$, which translates to a projected distance of
$\sim$ 10000 AU for the median distance of the exoplanet sample.  This
is of the order of magnitude of the canonical limit for gravitational
binding, although \citet{Pov1994} listed several companions with
separations larger than this.

Of the 131 systems, 82 had easily detectable proper motions and hence
were searched effectively for CPM companions, 7 had marginal proper
motions, and 42 systems had no detectable proper motions.  Of the 82
systems searched effectively, 15 definite CPM companions were
confirmed (one per system), and 67 had no CPM companions detected
within the search region outlined above.  However, in 12 (plus 3
candidates) of these 67 systems, close companions were identified by
other sources.  In 3 (plus 3 candidates) of the 49 marginal or
unsearched systems, companions were reported by other sources.  These
additional companions could not be detected by our method due to
saturation around the primary, and/or a short time baseline between the
DSS image pair.

\subsection {Washington Double Star Catalog (WDS)}

The WDS catalog\footnote{$http://ad.usno.navy.mil/wds/$} is the
world's most comprehensive database of multiple stars.  However, it is
a catalog of doubles, not binaries, so it explicitly contains an
unknown number of non-physical chance alignments.  Table~\ref{WDS-no}
lists 20 WDS entries that are not gravitationally bound to the
exoplanet host, but rather are field stars, listed in WDS ID sequence
(column 1).  The second column is the HD identifier of the star.  The
third column is the component suffix of the pair, as it appears in the
WDS catalog, for which position angle, separation and epoch of the
most recent observation are listed in columns four, five and six.  The
seventh column is the number of observations listed in the WDS.  Note
that a few of these ``companions'' have many observations, but they
are not true companions.  The eighth column identifies the specific
method used to refute the WDS entry.

Figure~\ref{WDS-CPM} shows an example for HD 9826.  The lines mark two
WDS entries that do not share the primary's high proper motion and
hence are background stars.  On the other hand, the known CPM
companion (marked by an arrow) is easily identifiable in these images.

\subsection {Hipparcos Catalog}
\label{sec:Hip}

As most of the exoplanet systems are close to the Sun (128 of the 131
are within 100 pc), the Hipparcos
Catalog\footnote{$http://www.rssd.esa.int/Hipparcos/HIPcatalogueSearch.html$}
provides fairly reliable distances and some photometric data for these
systems.  The catalog also notes some stellar companions, identified
by field H59 as component solutions ('C' flag), accelerated proper
motion ('G' flag), or orbital solutions ('O' flag).  In total,
Hipparcos identified stellar companions in nine exoplanet systems,
four each with 'C' and 'G' flags, and one with the 'O' flag.  Five of
the nine Hipparcos companions were independently confirmed, one (HD
38529c) is a close brown dwarf, and two (both 'G' flags) remain as
candidates.  The $\rho$ CrB system (HD 143761) has an 'O' flag, and
contains a companion that is a planet \citep{Noy1997,Zuc2001} or a
star \citep{Gat2001,Pou2001,Hal2003}, but not both.

\subsection {Catalog of Nearby Stars (CNS)}

Among our sample of 131 stars, 39 are listed in the CNS.  We reviewed
the earlier versions of the catalog \citep{CNS1969,CNS1979,CNS1991},
as well as the consolidated information on the
web\footnote{$http://www.ari.uni-heidelberg.de/aricns/$}.  The catalog
identifies any known companions, and lists separation, position angle
and references in the notes section.  Twelve stars from our sample
have companions listed in the CNS, and every one of them was confirmed by
other sources to be a true companion.

\subsection {Duquennoy \& Mayor}

The \citet{Duq1991} G Dwarf Survey specifically looked at multiplicity
among solar-type stars in the solar neighborhood using radial velocity
techniques.  This is an ideal reference for our sample because
searches for exoplanet systems have focused on such systems.
\citeauthor{Duq1991} identified target stars as single-line, double-line,
or line-width spectroscopic binaries, or spectroscopic binaries with
orbits.  Only three stars from our sample have companions listed in
this reference, and each of these was confirmed by other sources to be
a true companion.

\section {Photometric Distance Estimates for Companion Candidates}

In addition to the proper motion investigation, we collected archival
2MASS and SuperCOSMOS photometry as well as new CCD photometry that
allowed us to compute distance estimates to companion candidates, as
described below.  Table~\ref{Obs} summarizes the photometry data, as
well as the distance estimates computed.  The first column is the
star's name, and the second column contains the spectral type
identified as part of this work (see \S\ref{sec:SpecObs}).  The next
three columns are the $BRI$ plate magnitudes from SuperCOSMOS,
followed by the $VRI$ CCD magnitudes observed by us at the CTIO 0.9m
and 1.0m telescopes. The ninth column gives the number of observations
available for the $VRI$ photometry.  This is followed by 2MASS $JHK_S$
photometry.  The thirteenth, fourteenth and fifteenth columns are the
estimated plate photometric distance, total error of this estimate,
and the number of color relations used in computing this estimate.
The last three columns similarly list the CCD distance estimate, total
error and the number of color relations used.

\subsection {2MASS Coordinates \& Photometry}

We used the 2MASS web database, accessed via the Aladin interactive
sky atlas\footnote{$http://aladin.u-strasbg.fr/aladin.gml$}
\citep{Aladin} to obtain equinox 2000 coordinates for the companion
candidates, the epoch of observation, and $J$, $H$ and $K_S$
photometry.  The errors in $JHK_S$ were almost always less than 0.05
mag, and were typically 0.02--0.03 mag.  Notable exceptions are three
distant and faint refuted candidates listed in Table~\ref{Obs}, HD
33636 \#1 (errors of 0.14, 0.15, null at $JHK_S$ respectively), HD
41004 \#1 (0.05, 0.06, and 0.07 mag), and HD 72659 \#1 (0.05, 0.06 and
0.07 mag).

\subsection {SuperCOSMOS Plate Photometry and Distance Estimates}

We obtained optical plate photometry in $B_J$, $R_{59F}$ and $I_{IVN}$
bands (hereafter $BRI$) from the SuperCOSMOS Sky Survey (SSS) scans of
Schmidt survey plates \citep{Ham2001a}. The SSS plate photometry is
calibrated by means of a network of secondary standard star sequences
across the entire sky, with the calibration being propagated into
fields without standards by means of the ample overlap regions between
adjacent survey fields. The external accuracy of the calibrations is
$\pm$ 0.3 mag in individual passbands \citep{Ham2001b}; however the
internal accuracy in colors (e.g. $B-R$, $R-I$) is much better, being
typically 0.1 mag for well--exposed, uncrowded images. We used point
source photometric measures in all cases.

Photometric distance estimates were then derived using these SSS plate
magnitudes, combined with 2MASS $JHK_S$ by fitting various colors to
$M_{K_S}$--color relations from \citet{Ham2004}.  Results for 11
companion candidates are given in Table~\ref{Obs}.  Errors quoted from
this procedure include internal and external errors.  Internal errors
represent the standard deviation of distance estimates from the suite
of $M_{K_S}$--color relations.  External errors represent a measure of
the reliability of the relations for stars of known distance, which is
estimated to be 26\% in \citet{Ham2004}.

\subsection {CCD Photometry Observations and Distance Estimates}

Because of the relatively large photometric distance errors associated
with photographic plate photometry, we obtained optical CCD photometry
for one exoplanet host and 13 companion candidates (given in
Table~\ref{Obs}) in the $V_JR_{KC}I_{KC}$ bands (hereafter $VRI$)
using the Cerro Tololo Inter-American Observatory (CTIO) 0.9m and 1.0m
telescopes during observing runs in December 2003, June, September and
December 2004, August and December 2005, and March 2006 as part of the
SMARTS (Small and Moderate Aperture Research Telescope System)
Consortium.  For the 0.9m telescope, the central quarter of the 2048
$\times$ 2046 Tektronix CCD camera was used with the Tek 2 $VRI$
filter set.  For the 1.0m telescope, the Y4KCam CCD camera was used
with the Harris 1 4mts $VR$ and kc 1 4mts $I$ filter set.  Standard
stars from \citet{Gra1982}, \citet{Bes1990} and \citet{Lan1992} were
observed through a range of air masses each night to place measured
fluxes on the Johnson-Kron-Cousins $VRI$ system and to calculate
extinction corrections.

Data were reduced using IRAF via typical bias subtraction and dome
flat-fielding, using calibration frames taken at the beginning of each
night.  In general, a circular aperture 14$\arcsec$ in diameter was
used to determine stellar fluxes in order to match apertures used by
\citet{Lan1992} for the standard stars.  In cases of crowded fields,
an appropriate aperture 2$\arcsec$--12$\arcsec$ in diameter was used
to eliminate stray light from close sources and aperture corrections
were applied.  For one target (HD 169830B), we used Gaussian fitting
via an IDL program to the PSF tail of a bright nearby source to
eliminate its effects, and completed the photometry on the target
using IDL APER routine.  The same approach was performed on two of our
standard stars to correct for zero point difference between IDL and
IRAF magnitudes.  As discussed in \citet{Henry2004}, photometric
errors are typically $\pm$ 0.03 mag or less, which includes both
internal and external errors.  The only exceptions with larger errors
were distant and faint refuted candidates HD 33636 \#1 (errors of
0.06, 0.04, and 0.04 mag at $VRI$ respectively) and HD 72659 \#1
(0.10, 0.05, and 0.03 mag), new companion HD 188015B (0.05 and 0.04
mag at $R$ and $I$, respectively), and new candidate HD 169830B (0.12,
0.09, and 0.13 mag).  The errors for HD188015B and HD 169830B are high
due to the uncertainties introduced by the large aperture corrections
and, for HD 169830B, PSF fitting as well.

Photometric distances were obtained using the $VRI$ magnitudes along
with 2MASS $JHK_S$, and fitting various colors to $M_{K_S}$--color
relations from \citet{Henry2004}.  The results for these companion
candidates are given in the final three columns of
Table~\ref{Obs}. Errors quoted from this procedure include internal
and external errors.  Internal errors represent the standard deviation
of distance estimates from the suite of $M_{K_S}$--color relations.
External errors represent a measure of the reliability of the
relations for stars of known distance, which is estimated to be 15\%
in \citet{Henry2004}.

\section {Spectroscopic Observations}
\label{sec:SpecObs}

New spectra of nine companion candidates were obtained during
observing runs in October and December 2003, March and September 2004,
and January 2005 at the CTIO 1.5m telescope as part of the SMARTS
Consortium.  The R-C Spectrograph and Loral 1200 X 800 CCD detector
were used with grating \#32 in our red setup and \#09 in our blue
setup, which provided 8.6\AA~resolution and wavelength coverage from
6000-9500\AA~in the red and 3800-6800\AA~in the blue.  Data reduction
consisted of background subtraction, spectrum extraction, and
wavelength and flux calibrations in IRAF after standard bias
subtraction, flat fielding, and illumination corrections were applied.
Standard dome flats were used for flat fielding and calibration frames
were taken at the beginning of each night.  Fringing at wavelengths
longer than 7000\AA~is common in data from this spectrograph; however
it is typically removed fully by flat fielding, and no further steps
were needed to remove the fringes.  Spectral types for stars observed
in the red wavelength regime, listed in Table~\ref{Obs}, were assigned
using the ALLSTAR program as described in \citet{Henry2002}.  RECONS
types have been assigned using a set of standard comparison stars from
the RECONS database, a library of $\sim$ 500 M0.0V to M9.0V spectra.
Only rough spectral types were assigned based on our blue spectra by
comparing features in our spectra with standard stars in
\citet{Jac1984}.

\section {Results}

Table~\ref{results} is a compendium of the 30 exoplanet systems
confirmed to have two or more stellar components, listed in coordinate
sequence.  At the end of the table, six additional systems are listed
that may be stellar multiples, although these have not yet been
confirmed.  The first column lists a sequence number of the exoplanet
system matching the value plotted in Figure 5, and the second and
third columns list the HD name and an alternate name of the exoplanet
host and companion stars.  The fourth column lists stellar (A, B,
C...), or planetary components (b, c, d, ...).  The fifth column lists
the RA \& DEC of stars at epoch 2000, equinox 2000.  For stars listed
in Hipparcos (all primaries and a few companions), we used the
Hipparcos 1991.25 epoch coordinates and proper motions to compute the
coordinates listed. For fainter stars not observed by Hipparcos, we
used 2MASS coordinates at the epoch of observation, and converted the
coordinates to epoch 2000.0 using proper motions from SuperCOSMOS or
NLTT \citep{Luy1979}, if available.  When the proper motion of a
companion was not available, we used the primary's Hipparcos proper
motion.  In some instances, 2MASS coordinates were not available for
the companions, and in these instances, the coordinates of the
companions are not listed.  However, in all but three of these cases,
the separation and position angle of the companion from the primary
are listed in columns 10 and 11.  The three exceptional cases (one
confirmed and two candidates), where neither coordinates nor
separations from the primaries are known, are all Hipparcos 'G' flags,
and hence close astrometric binaries.  The sixth column lists the
trigonometric parallax from Hipparcos, in seconds of arc.  The seventh
and eighth columns list the distance, in parsecs, based on either
trigonometric parallax, if available (coded as 'T'), calculated CCD
photometric distance using relations from \citet{Henry2004} (coded as
'C'), or calculated plate magnitude distance from SuperCOSMOS using
relations from \citet{Ham2004} (coded as 'P').  If both plate and CCD
distance estimates are available, only the more reliable CCD distance
is listed.  The ninth column lists the spectral type from
\citet{Gra2003}, the planet discovery paper, or other references for
the primary, and from our spectroscopic observations or other
references for the companion.  The tenth and eleventh columns list the
angular separation (in seconds of arc) and position angle (in degrees)
of stellar companions with respect to the exoplanet host.  For
companions listed in WDS, these are typically the most recent entry in
WDS, otherwise they are the values listed in the companion discovery
paper.  For new companions, these astrometry values are our measurements
from our CTIO or the DSS images.  The twelfth column lists the
projected spatial separation (and is therefore a lower limit at the
epoch of plate observation) of companion stars with respect to their
primaries, in AU.  The thirteenth column gives the $M \sin{i}$ in
Jupiter masses for planets.  The fourteenth and fifteenth columns list
the a $\sin{i}$ (in AU) and eccentricity of the orbits.  The
sixteenth column specifies the sources used to detect the companion
stars.  The codes are as follows: 'P' represents a CPM detection using
the multi-epoch DSS images; 'W' represents a companion listing in the
WDS catalog; 'H' represents a Hipparcos catalog companion
identification; 'C' represents a companion identification in the CNS
catalog; 'D' represents a companion identification in \citet{Duq1991};
'I' represents confirmation via our recent $VRI$ images taken to
verify CPM; and 'O' represents that the companion was not found by any
of the above means, but reported in one or more refereed papers.
Finally, the seventeenth column lists relevant references relating to
stellar companions.  We have chosen not to list the individual planet
discovery papers as references, unless they identify a stellar
companion.

\subsection {Notes for Each Multiple System}
\label{sec:NotESys}

\subsubsection{New, Known, or Confirmed Companion Systems}
\label{sec:KnoCmp}

{\bf 1. HD 142:} This close binary (separation 5$\farcs$4) is listed
in WDS and CNS.  While this pair was first resolved at Harvard College
Observatory in 1894 \citep{Bai1900}, the separation and $\Delta$m
$\simeq$ 5 make this a difficult object. It was found at approximately
the same position six times from 1894 to 1928.  It then remained
unmeasured for 72 years until it became evident in 2MASS in 2000 at
approximately the same position angle.  Given the primary's $\mu$ =
0$\farcs$58 yr$^{-1}$ due east, and the long time lapse between the
1928 WDS observation and our image of 2004, a background star would
easily have been detected, but we found a blank field at its expected
position.  This system was mentioned in \citet{Low2002} as a single
planet in a multiple star system.

{\bf 2. HD 9826:} This CPM pair is clearly identified in DSS images,
but not listed in any of the other sources checked.  \citet{Low2002}
identified this as the first system discovered with multiple planets
and multiple stars.  It was also mentioned in \citet{Pat2002} and
\citet{Egg2004} as an exoplanet primary having a stellar companion.

{\bf 3. HD 11964:} This CPM pair is clearly identified in DSS images,
and listed in WDS and CNS.  \citet{All2000} listed this as a wide
binary system in a catalog of 122 binaries identified via CPM from a
sample of 1,200 high-velocity, metal-poor stars.  The primary's $\mu$
= 0\farcs441 yr$^{-1}$ at 236\sdeg~from Hipparcos, and the companion's
$\mu$ = 0\farcs444 yr$^{-1}$ at 236\sdeg~\citep{UCAC2}, a good match.
Our work is the first identification of this as a stellar companion to
a planetary system.

{\bf 4. HD 13445:} \citet{Els2001} reported the discovery of this
close companion (1\farcs72 $\pm$ 0\farcs2 separation) via AO imaging,
incorrectly identifying the companion as a T-dwarf based on its
colors.  The recent publication of \citet{Mug2005a} identified this
companion as a cool white dwarf based on its spectrum, claiming the
first white dwarf discovery in a planetary system.  However, HD 147513
was in fact the first white dwarf discovery in a planetary system,
reported by \citet{May2004}.  There are now two known systems with
evidence of planets surviving the post-main-sequence evolution of a
stellar companion, with this one being the closest known white dwarf
companion to an exoplanet host (at a projected separation of just 21
AU -- similar to Sun-Uranus distance).

{\bf 5. HD 19994:} WDS lists 14 observations for this companion.
This pair was first resolved by Admiral Smyth in 1836 with a 6 inch
refractor \citep{Smy1844}. It has been resolved fifteen times since
then, most recently by \citet{Hale1994} who also calculated a 1420 yr
orbit for this pair. While there is some hint of curvilinear motion in
the data, the orbit is certainly preliminary.  This companion is also
listed in CNS and \citet{Duq1991}.  Several references have identified
this as a stellar companion to a planetary system
\citep{Low2002,May2004,Egg2004,Udry2004}.

{\bf 6. HD 27442:} WDS and CNS list this companion at 13\farcs7
separation at 34\sdeg. It was first resolved in 1930 by
\citet{Jes1955}, and measured again by \citet{Hold1966} almost 35
years later at approximately the same position.  Our short-exposure
$VRI$ images taken at CTIO in September 2004 identified a source about
13$\arcsec$ away at 34\sdeg, consistent with the observations of
almost 75 years ago.  Given the primary's $\mu$ = 0\farcs175
yr$^{-1}$, this can be confirmed as a companion.  Our work is the
first identification of this as a stellar companion to a planetary
system.

{\bf 7. HD 38529:} This CPM pair was discovered by us using DSS
images.  The primary's $\mu$ = 0\farcs163 yr$^{-1}$ at 209\sdeg~from
Hipparcos, and the companion's $\mu$ = 0\farcs162 yr$^{-1}$ at
204\sdeg~from \citet{Lep2005} and 0\farcs158 yr$^{-1}$ at
208\sdeg~from SuperCOSMOS.  Figure~\ref{new38529} includes two DSS
images showing the primary and the companion.  Our CCD photometric
distance estimate of 28.7 $\pm$ 4.8 pc is consistent with our spectral
identification of M3.0V and matches the primary's distance of 42 pc
within 3$\sigma$.  At our request, spectroscopic observations of the
companion were obtained by G. Fritz Benedict in February 2004 using
the McDonald Observatory 2.1m telescope and Sandiford Cassegrain
echelle spectrograph \citep{McC1993}.  The data were reduced and 1-D
spectra were extracted using the standard IRAF \textit{echelle}
package tools.  The radial velocity was determined by
cross-correlating the spectra of the star with that of an M2 dwarf (GJ
623) template using the IRAF task \textit{fxcor}.  The adopted radial
velocity for the GJ 623 primary (it is a binary) was $-$29.2 km
s$^{-1}$, given the orbital phase at which the template was secured
and a systemic velocity of $-$27.5 km s$^{-1}$, from \citet{Mar1989}.
HD 38529B's radial velocity was measured to be 26.26 $\pm$ 0.10 km
s$^{-1}$.  This is roughly consistent with the primary's radial
velocity of 30.21 km s$^{-1}$ \citep{Nid2002}, and the odds of two
unassociated stars having such similar velocities are low.  However,
discrepancies in radial velocities and photometric distances could
indicate that the new companion is a double.  The projected separation
of the primary to the new companion(s) is $\sim$ 12000 AU, which is
extreme for a gravitationally bound system, although \citet{Pov1994}
listed a few wide binaries with even greater separations.  This
primary also has a Hipparcos 'G' flag listing, which was recently used
by \citet{Ref2006} to conclude that the sub-stellar companion ``c'' is
actually a brown dwarf of mass 37$_{-19}^{+36}$ \mjup.

{\bf 8. HD 41004:} A companion is listed in WDS and annotated in
Hipparcos with a 'C' flag, indicating a linear relative motion of
components, implying either an orbital period that is several times
the length of the Hipparcos observing interval (3.3 years), or stars
that are not physically linked.  At a separation of 0\farcs5 and a
$\Delta$m = 3.67 (from Hipparcos), the identification of a close
companion is difficult, but there are other such Hipparcos
observations (similar separation and $\Delta$m) that were
independently confirmed.  For example, T.J.J. See measured a close
large $\Delta$m pair, known as SEE 510 (HIP 86228), with the Lowell 24
inch telescope in 1896 \citep{See1896}. This pair, lost for nearly 100
years, was recovered by Hipparcos at about the same position
(0\farcs2, $\Delta$m = 1.8).  While SEE 510 isn't morphologically
identical to HD 41004, we believe that it is comparably difficult, and
so we accept the Hipparcos identification of a companion to HD 41004.
This system was mentioned in \citet{Egg2004} as a stellar companion in
an exoplanet system.  Further, \citet{Zuc2003} listed the radial
velocity for the primary as 42.5 $\pm$ 0.01 km s$^{-1}$, and found the
companion to be a double, with a velocity range of 34--48 km s$^{-1}$
($\pm$ 0.56 km s$^{-1}$) over 103 observations.  They derived an
orbital solution for the BC pair, concluding that the orbit is nearly
circular with a $\sin{i}$ = 0.016 AU, and that the low mass companion
has a minimum mass of 18.4 \mjup.  \citet{Zuc2004} derived orbital
elements of the possible M dwarf--brown dwarf pair and concluded that
this is a unique system with each stellar component of a visual binary
having a low mass companion in orbit around it --- one a planet, and
the other a possible brown dwarf.  Note that the projected separation
between A and B is just 22 AU, similar to the separation of the Sun
and Uranus.

{\bf 9. HD 40979:} This CPM pair is clearly identified in DSS images.
The primary is 33 pc away with $\mu$ = 0\farcs179 yr$^{-1}$ at
148\sdeg~(from Hipparcos).  The companion, BD+44 1351, has a very
similar $\mu$ = 0\farcs179 yr$^{-1}$ at 148\sdeg~from \citet{Lep2005}
and 0\farcs180 yr$^{-1}$ at 148\sdeg~from \citet{Hog1998}.
\citet{Hal1986} identified this CPM pair, listing the companion as a
K5 star.  \citet{Egg2004} identified this as a stellar companion to a
planetary system, noting that physical association of this pair has
been confirmed via radial velocity measurements.  However, our plate
photometric distance estimate to the companion is 15.2 $\pm$ 4.0 pc
(based on only 3 colors), not a very good match with the primary,
although the error is large.  This discrepancy could be due to the
poor quality of the photometric distance estimate (due to the blue
colors of the companion) or perhaps because the companion is an
unresolved double.

{\bf 10. HD 46375:} WDS lists this 9\farcs4 separation companion at
308\sdeg.  We took short exposure frames at CTIO in September 2004,
which identified a companion at a separation of 10$\arcsec$ at
310\sdeg, consistent with the WDS observation.  The first published
resolution of this pair made by \citet{Sou1985} in 1984 has also been
confirmed by 2MASS images.  Re-analysis of Astrographic Catalogue data
\citep{Urb1998} has added an observation at about the same secondary
position in 1932, thereby confirming that it has the same proper
motion.  Our CCD photometric distance estimate of 26.4 $\pm$ 6.0 pc is
within 2$\sigma$ of the primary's distance of 33.4 pc from Hipparcos.
We therefore conclude that this is a physical pair.  This work is the
first identification of this as a stellar companion to a planetary
system.

{\bf 11. HD 75289:} This CPM candidate was detected by
\citet{Mug2004b} and confirmed by their photometry and spectroscopy.
While the companion is seen in the epoch-2 DSS image, CPM could not be
established by our method due to saturation of the region around
around the primary in the epoch-1 image.

{\bf 12. HD 75732:} This CPM pair is easily identified in DSS images,
and matches entries in WDS, CNS and \citet{Duq1991}.  The primary has
$\mu$ = 0\farcs539 yr$^{-1}$ at 244\sdeg~and $\pi$ = 0\farcs07980
$\pm$ 0\farcs00084, from Hipparcos.  Our CCD photometric distance
estimate to the companion is 8.7 $\pm$ 1.4 pc, a match within
3$\sigma$.  The companion's $\mu$ = 0\farcs540 yr$^{-1}$ at
244\sdeg~and $\pi$ = 0\farcs0768 $\pm$ 0\farcs0024 from the Yale
Parallax Catalog \citep{vanAlt1995} and 0\farcs0757 $\pm$ 0\farcs0027
from \citet{Dahn1988} are all consistent with the primary's.  This
system is listed in \citet{Egg2004} as a stellar companion to a
planetary system.  The primary star, more commonly known as 55 Cnc,
has four reported planets, so this system is the most extensive solar
system with a stellar companion, which is at a projected distance of
more than 1000 AU. The discrepancy in photometric distance could hint
that the companion is an unresolved double.

{\bf 13. HD 80606:} This CPM pair is easily identified in DSS images,
and matches entries in WDS and Hipparcos.  The primary's $\mu$ =
0\farcs047 yr$^{-1}$ at 82\sdeg~and $\pi$ = 0\farcs01713 $\pm$
0\farcs00577, from Hipparcos.  The parallax has a large error due to
the close companion.  The companion is HD 80607, spectral type G5,
$\mu$ = 0\farcs043 yr$^{-1}$ at 79\sdeg, and Hipparcos lists an
identical parallax.  This companion was listed by \citet{Egg2004} as a
stellar companion to a planetary system.

{\bf 14. HD 89744:} This companion was reported as a candidate by
\citet{Wil2001} based on spectroscopic observations, and they
identified it as a massive brown dwarf of spectral type L0V.
Companionship was subsequently confirmed astrometrically by
\citet{Mug2004a}.  This faint companion is not seen in the DSS images.

{\bf 15. HD 99492:} This CPM pair is easily identified in DSS images,
and matches entries in WDS, Hipparcos and CNS.  Component B (the
exoplanet host) has $\mu$ = 0\farcs755 yr$^{-1}$ at 285\sdeg~and $\pi$
= 0\farcs05559 $\pm$ 0\farcs00331, from Hipparcos.  Component A is HD
99491 with spectral type K0IV, $\mu$ = 0\farcs749 yr$^{-1}$ at 284\sdeg,
and $\pi$ = 0\farcs05659 $\pm$ 0\farcs00140, from Hipparcos.  These
match HD 99492's values well, and confirm the pair as physical.

{\bf 16. HD 114729:} This CPM candidate was detected recently by
\citet{Mug2005b} and confirmed by their photometry and spectroscopy.
It could not be detected using DSS frames due to saturation of the
region around around the primary.  

{\bf 17. HD 114762:} This close companion was discovered using
high-resolution imaging \citep{Pat2002}. It was also mentioned by
\citet{Egg2004} as a stellar companion to a planetary system.  The
``planet'', with $M \sin{i}$ = 11.0 \mjup~may in fact be a star 
in a low inclination orbit \citep{Coc1991,Fis2005}.

{\bf 18. HD 120136:} This close companion is listed in WDS (53
observations), CNS and in \citet{Duq1991}.  The primary's $\mu$ =
0\farcs483 yr$^{-1}$ at 276\sdeg~from Hipparcos.  CNS lists the
companion as GJ 527B, and SIMBAD gives its $\mu$ = 0\farcs480
yr$^{-1}$ at 274\sdeg, a good match to the primary's.  This system has
been recognized as a stellar companion to an exoplanet system
\citep{Pat2002,Egg2004}.

{\bf 19. HD 142022:} This CPM pair (GJ 606.1AB) is easily identified
in DSS images, and matches entries in WDS and CNS.  The companion's
spectral type is K7V.  The NLTT catalog lists identical $\mu$ for both
components = 0\farcs320 yr$^{-1}$ at 269\sdeg~\citep{Luy1979}.

{\bf 20. HD 147513:} This companion is listed in CNS and was the first
white dwarf found in an exoplanet system \citep{May2004}.  The
primary's $\mu$ = 0\farcs073 yr$^{-1}$ at 87\sdeg~and $\pi$ =
0\farcs07769 $\pm$ 0\farcs00086, from Hipparcos.  The companion is HIP
80300, type DA2 \citep{Weg1973}, with matching Hipparcos values of
$\mu$ = 0\farcs076 yr$^{-1}$ at 90\sdeg~and $\pi$ = 0\farcs07804 $\pm$
0\farcs00240.

{\bf 21. HD 178911:} This is a triple star system with one known
planet.  The wide CPM pair (AC-B) is clearly seen in DSS images.  The
6.3 \mjup~planet orbits HD 178911B, while HD 178911AC is a close
binary, first resolved by \citet{McAl1987a} with the
Canada-France-Hawaii Telescope (CFHT).  This pair has since been
resolved ten more times, most recently with the 6m telescope of the
Special Astrophysical Observatory in Zelenchuk in 1999
\citep{Bal2004}.  \citet{Hart2000} present an orbital solution with a
3.5 year period based on speckle observations and \citet{Tok2000}
present a full orbital solution using spectroscopic and
interferometric data.  The multiplicity of this system has been
previously identified \citep{Zuc2002,Egg2004}.  From Hipparcos, HD
178911AC's $\mu$ = 0\farcs200 yr$^{-1}$ at 14\sdeg~and $\pi$ =
0\farcs02042 $\pm$ 0\farcs00157 and the companion's $\mu$ = 0\farcs203
yr$^{-1}$ at 19\sdeg~and $\pi$ = 0\farcs02140 $\pm$ 0\farcs00495, a
match within the errors, confirming a physical association.

{\bf 22. HD 186427:} This is a triple star system with one known
planet. The wide CPM pair (AC-B) is clearly seen in DSS images.  The
planet orbits 16 Cyg B (HD 186427), while 16 Cyg A (HD 186408) is a
close binary, first resolved by \citet{Turn2001} with the AO system on
the Hooker 100'' telescope, and independently confirmed by IR imaging
by \citet{Pat2002} with the Keck 10m and Lick 3m.  In the five total
observations, the position of the secondary has not changed much.
However, they span less than two years of time and little motion would
be expected at a projected separation of 73 AU.  The multiplicity of
this system has been previously identified
\citep{Pat2002,Low2002,Egg2004}. From Hipparcos, 16 Cyg A's $\mu$ =
0\farcs217 yr$^{-1}$ at 223\sdeg~and $\pi$ = 0\farcs04625 $\pm$
0\farcs00050 and the planet host's $\mu$ = 0\farcs212 yr$^{-1}$ at
220\sdeg~and $\pi$ = and 0\farcs04670 $\pm$ 0\farcs00052, a match
within the errors, confirming a physical association.

{\bf 23. HD 188015:} This new companion was detected by us as a CPM
candidate and confirmed via CCD photometry.  The primary's $\pi$ =
0\farcs01900 $\pm$ 0\farcs00095 and $\mu$ = 0\farcs106 yr$^{-1}$ at
149\sdeg, from Hipparcos.  The companion, 13$\arcsec$ away from the
primary at 85\sdeg, does not have proper motion listed in SuperCOSMOS
or NLTT, but our CCD photometric distance of 46.9 $\pm$ 9.5 pc matches
the primary's distance within 1$\sigma$, and hence confirms this as a
companion.  Figure~\ref{new188015} includes two DSS images showing the
primary and the companion.

{\bf 24. HD 190360:} This CPM pair is easily identified in DSS images,
and matches entries in WDS and CNS.  The primary is GJ 777A with
spectral type G7IV-V and $\mu$ = 0\farcs861 yr$^{-1}$ at
127\sdeg~from Hipparcos.  The companion is GJ 777B with spectral type
M4.5V and $\mu$ = 0\farcs860 yr$^{-1}$ at 127\sdeg~\citep{Lep2005}.
Our plate photometric distance estimate of 18.5 $\pm$ 6.2 pc is a good
match with the primary's trigonometric parallax distance of 15.9 pc.
This system has been recognized as a binary, and as an exoplanet
primary with a stellar companion \citep{All2000,Naef2003,Egg2004}.

{\bf 25. HD 195019:} WDS is the only source listing this close binary
at a separation of 3\farcs5 at 330\sdeg.  The close pair, first
resolved by \citet{Hou1887} with an 18 inch refractor, has moved 7
degrees in position angle and closed in from 4\farcs5 to 3\farcs5 in
separation in 12 observations over 107 years.  This transition has not
been smooth, no doubt due to $\Delta$m = 4, making observations a
challenge.  The typical measurement errors of micrometry coupled with
slow motion makes characterization difficult.  It was identified as a
binary in \citet{All2000}, and recognized as a stellar companion to an
exoplanet system in \citet{Egg2004}.

{\bf 26. HD 196050:} This CPM candidate was detected recently by
\citet{Mug2005b} and confirmed by their photometry and spectroscopy.
It could not be detected using DSS frames due to saturation of the
region around around the primary.

{\bf 27. HD 213240:} This CPM pair was identified by us using DSS
images.  The primary's $\mu$ = 0\farcs236 yr$^{-1}$ at 215\sdeg~and
$\pi$ = 0\farcs02454 $\pm$ 0\farcs00081, from Hipparcos.  The
companion's $\mu$ = 0\farcs229 yr$^{-1}$ at 214\sdeg~from SuperCOSMOS
is a good match.  Our CCD photometric distance of 41.8 $\pm$ 6.5 pc is
consistent with our spectral type identification of M5.0V, and is a
good match to the primary's trigonometric parallax distance of 40.8
pc.  This new companion identification in an exoplanet system was
recently reported by \citet{Mug2005b} during the writing of this
paper.

{\bf 28. HD 219449:} A CPM companion is easily detected in the DSS
images, and is matched by WDS and CNS entries.  WDS lists the
secondary as a tight binary (0\farcs4 separation at 101\sdeg).  The
primary's $\mu$ = 0\farcs369 yr$^{-1}$ at 93\sdeg~and $\pi$ =
0\farcs02197 $\pm$ 0\farcs00089, from Hipparcos.  The companion binary
has $\mu$ = 0\farcs377 yr$^{-1}$ at 91\sdeg~from NLTT and 0\farcs385
yr$^{-1}$ at 96\sdeg~from \citet{UCAC2}, both good matches to the
primary.  NLTT also lists the companion's spectral type as K8V.  Our
CCD photometric distance of 29.9 $\pm$ 4.7 pc is for the BC pair, and
we predict an actual distance of 42.4 pc (assuming identical spectral
types), which is a good match to the primary (45.5 pc).  Radial
velocities from \citet{Wil1953} are -26.4 $\pm$ 0.9 km s$^{-1}$ for
the primary and -25 $\pm$ 5 km s$^{-1}$ for the secondary, also a
match within the errors.  Our approximate spectral identification as
an early K type is consistent with the photometric distances.  This
work recognizes, for the fist time, that this exoplanet system resides
in a triple star system.

{\bf 29. HD 222404:} This companion is listed in Hipparcos with a 'G'
flag, indicating a close astrometric binary.  While some speckle
searches have failed to detect a companion (e.g. \citealt{Msn2001}),
the companion has been detected via radial velocity efforts and
identified as a stellar companion in an exoplanet system
\citep{Cam1988,Gri2002,Egg2004}.  The semi-major axes of the planet
and stellar companions with respect to the primary place them at
Sun-Mars and Sun-Uranus separations, respectively.

{\bf 30. HD 222582:} This CPM pair is easily seen in DSS images, and
is listed in the WDS.  The primary's $\mu$ = 0\farcs183 yr$^{-1}$ at
233\sdeg~and $\pi$ = 0\farcs02384 $\pm$ 0\farcs00111, from Hipparcos.
The secondary's $\mu$ = 0\farcs180 yr$^{-1}$ at 231\sdeg~from NLTT,
0\farcs186 yr$^{-1}$ at 230\sdeg~from SuperCOSMOS and 0\farcs187
yr$^{-1}$ at 232\sdeg~from \citet{UCAC2} are all good matches to the
primary.  Our CCD photometric distance of 32.1 $\pm$ 5.0 pc matches
the primary's distance of 42.0 pc within 2$\sigma$.  Our spectral type
of M3.5V is consistent with the photometric distance estimates.  This
pair, resolved by Luyten in 1960, was noted to have a common proper
motion.  This work confirms the gravitational relationship via CPM,
photometry, and spectroscopy and is the first identification of this
stellar companion to an exoplanet system.

\subsubsection{Candidate Companion Systems}

{\bf 31. HD 8673:} WDS is the only source listing a close companion,
at 0\farcs1 separation.  Resolved by Mason et al. (2005) as part of a
survey of nearby G Dwarfs for duplicity, this unpublished observation
has yet to be confirmed. The projected stellar separation of 3.8 AU is
just over twice the planet/brown-dwarf projected separation of 1.6 AU
and dynamical instability is likely.  Alternatively, given the large
$M \sin{i}$ = 14 \mjup~for the ``planet'', it is possible that it is
actually a star in a near face-on orbit (i $\le$ 10 degrees), and that
the radial velocity and speckle observations are of the same object.

{\bf 32. HD 16141:} This CPM candidate was recently detected by
\citet{Mug2005b} at a separation of 6\farcs2, and they plan follow-up
observations to confirm it.  We could not detect the companion using
DSS frames due to saturation of the region around around the primary.

{\bf 33. HD 111232:} This companion is mentioned only in Hipparcos,
and is listed with a 'G' flag, indicating that the proper motion was
best fit with with higher-order terms.  \citet{Msn1998} conducted a
specific search for a companion using optical speckle, but did not
find any.  Their effort should have picked up companions with
$\Delta$V $\sim$ 3 and separations 0\farcs035 -- 1\farcs08.  

{\bf 34. HD 150706:} This companion is mentioned only in Hipparcos,
and is listed with a 'G' flag, indicating that the proper motion was
best fit with with higher-order terms.  \citet{Hal2003} reported this
as a single star based on two CORAVEL radial velocity surveys that
yielded statistical properties of main-sequence binaries with spectral
types F7 to K and with periods up to 10 years.

{\bf 35. HD 169830:} A candidate companion was detected by Kevin Apps
as a close 2MASS source with 11$\arcsec$ separation at
265\sdeg~(private communication).  Our CCD photometric distance
estimate for the companion is 29 $\pm$ 23 pc, consistent with the
primary's distance of 36 pc, but the large error in our estimate
prevents confirmation.  The large error is likely due to the
uncertainty in our and 2MASS photometry, caused by the close, bright
primary, and the proximity of the companion to the primary's
diffraction spike.  While 2MASS lists errors of 0.04 mag for $JHK_S$,
it notes that the photometry is contaminated by a nearby bright
source. Also, the $J$ magnitude from DENIS is 0.36 magnitudes brighter
than the 2MASS value, indicating a larger uncertainty.  The low proper
motion (0\farcs015 yr$^{-1}$) of the primary prevents confirmation via
CPM.  While we believe that the evidence strongly indicates this as a
true companion, we can not confirm it until we obtain a spectrum or
other conclusive evidence.

{\bf 36. HD 217107:} Only WDS lists this close companion with 0\farcs3
separation at 156\sdeg.  Proper motion of the primary is not
detectable in DSS images.  This pair has been resolved only twice
\citep{McAl1987b,Msn1999} fifteen years apart, and the lack of
additional resolutions of this bright pair seems to indicate that a
large magnitude difference may be preventing additional detections.
Given the two reported planets with a $\sin{i}$ = 0.1 AU and 4.3 AU,
this companion at a projected separation of just 6 AU would likely
induce dynamical instability.  Explanations for this include the
possibility that this is an unrelated star with a chance
alignment, and/or that the wider ``planet'' is actually a stellar
companion with a highly inclined orbit.

\subsection{Refuted Candidates -- CPM alone does not confirm companionship}

As CPM is often used to detect gravitationally-bound
companions, we list here five exceptions that, upon follow-up
analyses, turned out to be unrelated field stars rather than true
companions.  In three of these instances (HD 33636, HD 41004 and HD
72659), we found proper motions in DSS plates to be an acceptable
match by eye, but photometric distances indicated that each candidate
was a distant field star.  In the cases of BD$-$10 3166 and HD 114783,
photometric distances did not provide a conclusive answer, but
plotting these on a $M_V$ versus $B-V$ curve of a sample of Hipparcos
stars allowed us to refute them.

{\bf BD$-$10 3166} is the only exoplanet primary without a Hipparcos
parallax.  We derived a CCD photometric distance of 66.8 $\pm$ 10.0
pc, but that is based on just one color because the object is on the
blue end of the $M_{K_S}$--color relations described in
\citet{Henry2004}.  The companion candidate, LP 731-076 is 17$\arcsec$
from the primary at 217\sdeg~(in the $DSS1$, epoch 1983.29 image), and
appears to have a matching proper motion.  The two stars were
identified by \citet{Luy1978} as a CPM pair, and recently recovered in
SuperCOSMOS data by Richard Jaworski (private communication).  In
SuperCOSMOS, the primary's $\mu$ = 0\farcs189 yr$^{-1}$ at
252\sdeg~and the candidate's $\mu$ = 0\farcs202 yr$^{-1}$ at 242\sdeg.
The candidate has a published photometric distance of 11.6 $\pm$ 0.8
pc \citep{Reid2002}, which is consistent with our photometric distance
estimate of 12.5 $\pm$ 2.0 pc and our spectral type listed in
Table~\ref{Obs}.  In order to get a better distance estimate to the
primary, we plotted 285 stars from Hipparcos on a $M_V$ versus $B-V$
diagram.  The stars were selected based on distance (parallax greater
than 0\farcs05), quality of parallax (error less than 10\%),
luminosity class (main sequence only), and $B-V$ value of greater than
0.5.  Fitting the primary's $B-V$ of 0.84 from \citet{Ryan1992} to the
least-squares-fit curve through the Hipparcos data yields a distance
estimate of 68 pc, consistent with our photometric distance estimate,
and too large to be associated with the candidate companion.  This is
an interesting example of a close (17$\arcsec$ separation) CPM pair
for which distance estimates to both components are of the same order
of magnitude, but the components seem to be unrelated.

{\bf HD 33636} has $\mu$ = 0\farcs227 yr$^{-1}$ at 127\sdeg~and $\pi$
= 0\farcs03485 $\pm$ 0\farcs00133 (29 pc) from Hipparcos.  The faint
CPM candidate at a separation of 220$\arcsec$ at 250\sdeg~(in the DSS
$POSS2/UKSTU Red$, epoch 1990.81 image) was refuted by us after
obtaining a CCD photometric distance of 739 $\pm$ 162 pc.  Our
spectrum, although noisy, allows us to estimate the spectral type to
be M1.0V, which indicates a large distance consistent with the
photometric estimate.

{\bf HD 41004} has $\mu$ = 0\farcs078 yr$^{-1}$ at 327\sdeg~and $\pi$
= 0\farcs02324 $\pm$ 0\farcs00102 (43 pc), from Hipparcos.  The faint
CPM candidate at a separation of 145$\arcsec$ at 335\sdeg~(in the DSS
$POSS2/UKSTU Red$, epoch 1993.96 image) was refuted by us after
obtaining a CCD photometric distance of 557 $\pm$ 103 pc.  We estimate
the spectral type to be M0.5, although the luminosity class is
uncertain -- it could be a dwarf or a sub-dwarf.  The candidate's
$\mu$ = 0\farcs046 yr$^{-1}$ at 6\sdeg~from SuperCOSMOS is not a good
match to the primary.

{\bf HD 72659} has $\mu$ = 0\farcs150 yr$^{-1}$ at 229\sdeg~and $\pi$
= 0\farcs01947 $\pm$ 0\farcs00103 (51 pc), from Hipparcos.  The
candidate, at a separation of 195$\arcsec$ at 165\sdeg~(in the DSS
$POSS2/UKSTU Red$, epoch 1992.03 image), was refuted by us after
obtaining a CCD photometric distance of 369 $\pm$ 99 pc.  Our spectral
identification as M3.0V is consistent with this photometric distance.
SuperCOSMOS lists the CPM candidate's $\mu$ = 0\farcs066 yr$^{-1}$ at
199\sdeg, showing that proper motion is not a good match.

{\bf HD 114783} is another CPM pair that looks like it may be
physical, but is not.  From SuperCOSMOS, the primary has $\mu$ =
0\farcs179 yr$^{-1}$ at 280\sdeg~and the candidate companion (at a
separation of 240$\arcsec$ at 47\sdeg~in the DSS $POSS2/UKSTU Red$,
epoch 1996.23 image) has $\mu$ = 0\farcs184 yr$^{-1}$ at 281\sdeg.
The primary's distance from the Hipparcos parallax is 20.4 pc.  Our
plate photometric distance estimate for the companion is 20.2 $\pm$
5.2 pc based on only 3 colors.  However, using CCD photometry, we get
a distance of 54.0 $\pm$ 9.3 pc, based on only 2 colors.  The
candidate companion is CCDM J13129-0213AB, a binary (listed in the WDS
with a separation of 2\farcs0 at 28\sdeg), and hence, its actual
distance is greater than photometrically indicated.  We plotted the
primary on the $M_V$ versus $B-V$ diagram using Hipparcos data as
described above, and it falls close to the main sequence fit,
indicating that it is likely a single star.  The candidate companion,
based on its $B-V$ of 1.10 yields a distance of 36 pc, using the
Hipparcos plot, but its actual distance will be greater because it is
a binary.  Our spectra for the two stars show very similar absorption
lines, although the continuum seems to indicate that the candidate
companion is slightly redder.  Given that the spectral types are
close, and that the candidate companion is a binary while the primary
appears to be single, we can only explain the large $\Delta$V (primary
$V$ = 7.56, and candidate companion $V$ = 9.78) by adopting
significantly different distances to the two stars.  Hence, we
conclude that this is not a gravitationally bound pair, despite the
compelling proper motion match.

\section {Discussion}
\label{sec:Discussion}

Our findings indicate that 30 (23\%) of the 131 exoplanet systems have
confirmed stellar companions, and 6 more (5\%) have candidate
companions.  Given the constraints of our search -- any new companions
we detected had to be widely-separated from primaries with high proper
motion -- these numbers should be regarded as lower limits.  This
point is confirmed by a recent paper, \citet{Mug2005b}, which reported
four new companions in exoplanet systems, of which we had
independently identified only one (HD 213240B).  Several interesting
properties are revealed by this comprehensive assessment.

Three of the exoplanet systems (HD 178911, 16 Cyg B, and HD 219449)
are stellar triples, and are arranged similarly --- a single planet
orbits close to one star and there is a distant, tight binary.  In
each system, the three stars are all of the same spectral class (G for
HD 178911 and 16 Cyg, and K for HD 219449).  We find it curious that
all three triple systems contain stars of comparable mass
(i.e. systems such as a G-dwarf exoplanet host with a M-dwarf binary
are not seen).  Could this be due to a selection effect (i.e. faint
companions are not as well studied for multiplicity) or does this say
something about the angular momentum distribution in star forming
regions?  Only a comprehensive survey of all companions for duplicity
can lead us to an answer.

It is interesting to note that recent exoplanet discoveries are
predominantly found in single star systems.  Of the first 102
radial-velocity-detected exoplanet systems, 26 (26\%) have confirmed
stellar companions.  In contrast, only 4 (14\%) of the latest 29
systems have confirmed stellar companions.  Even though we are dealing
with small number statistics, we believe that this change is
significant and worthy of further examination.  Our first inclination
was that recent planet detections are at larger projected semi-major
axes, and hence favor single systems because stellar companions would
have to be even farther out to provide the uncorrupted ``single''
systems sought by radial velocity programs.  However, we found no
correlation between the timing of exoplanet reporting and its
projected semi-major axis.  So, we are not able to explain this
curiosity at this point, and simply identify it for further
examination.

Exoplanet hosts are deficient in having stellar companions when
compared to a sample of field stars. Our updated results for stellar
counts in the exoplanet sample yield a single:double:triple:quadruple
percentage of 79:21:2:0 for confirmed systems, and 72:24:4:0
considering candidates.  While these are lower limits for
multiplicity, they are significantly lower than the \citet{Duq1991}
results of 57:38:4:1 for multiples with orbits, and 51:40:7:2
considering candidates.  This is certainly due in part to the fact
that planet searches specifically exclude known close binaries from
their samples (e.g. \citealt{Vogt2000}), and further eliminate any new
binaries detected via radial velocity.  We currently do not have
enough detailed information about the exoplanet search target
selection process to say whether the different multiplicity ratios are
entirely due to selection effects, or is indicative of planetary disk
instability and reduced planet formation in binary star systems.

\subsection {Planetary \& Stellar Orbits in Multiple Star Systems}

Figure~\ref{Orbits} shows the a $\sin{i}$ of planetary companions and
projected separations of the stellar companions for the 30 confirmed
exoplanets that reside in multiple star systems.  The Y-axis shows the
sequence number of the exoplanet system as listed in column 1 of
Table~\ref{results}.  The figure clearly indicates the presence of
separate planetary and stellar orbit regimes for the data currently
available.  All planets are within 6 AU, and all stars are at a
projected separation of greater than 20 AU from the exoplanet host.
Note that all points in the figure can potentially move right because
(1) planets are plotted at a separation of a $\sin{i}$, and (2) stars
are plotted based on their projected separations (although a few could
move left if they have been caught near apastron in their orbits).
The continued search for wider orbit planets will answer the question
of whether this is simply due to selection effect or if this says
something significant about planetary disk truncation in multiple star
systems.

55 Cnc (HD 75732), an extensive extrasolar system with four reported
planets, has the widest projected planetary orbit with an a $\sin{i}$
of 5.3 AU.  It is noteworthy that such an extensive exoplanet system
also has a stellar companion, at a projected separation of 1050
AU. This provides direct evidence of the stability of protoplanetary
disks in multiple star systems such as to allow formation and
sustenance of multiple planets, at least as long as the separation
between the stars is sufficiently large.  This system can also provide
an observational constraint for evaluating theoretical models of disk
stability and solar system evolution.

The smallest projected separation for a stellar companion is 21 AU for
GJ 86, closely followed by 22 AU for HD 41004 and $\gamma$ Cep.  Each
system has only one reported planet, with a $\sin{i}$ of 0.1 AU, 1.3
AU and 2.0 AU, respectively.  This may be evidence that a sufficiently
close stellar companion will disrupt the protoplanetary disk,
truncating planet formation at a few AU from the primary.

Every exoplanet system so far discovered in a multiple star system has
an S-type (Satellite-type) orbit, where the planet orbits one of the
stars.  This is not surprising because current radial velocity
searches for exoplanets exclude close binaries
(e.g. \citealt{Vogt2000}).  While the formation and stability of
planets in P-type (Planet-type) orbits, where a planet orbits the
center-of-mass of a binary or multiple star system in a circumbinary
configuration, has been theoretically demonstrated
\citep{Hol1999,Boss2005,Mus2005}, it has not yet been observationally
supported. However, \citet{Cor2005} have raised the interesting
possibility that the 2.4 \mjup~outer planet around HD 202206 may
in-fact have formed in a circumbinary disk around the primary and the
closer 17 \mjup~minimum mass object.

Several studies have investigated the theoretical stability of
planetary orbits in multiple star systems
(e.g. \citealt{Hol1999,Ben2003}), deriving ratios of orbital
semi-major axes of the planet and stellar companions for various
values of mass-ratio and eccentricity of the stellar orbits.  Our work
provides observational constraints based on all known exoplanets in
multiple star systems.  Of the 30 confirmed exoplanets in multiple
star systems, only three have a ratio of stellar to planetary
projected separation of less than 100.  The lowest ratio is 11, for
$\gamma$ Cep (HD 222404).  Although numerical simulations demonstrate
the stability of orbits for much smaller separation ratios (e.g. for
m$_2$/(m$_1$ + m$_2$) = 0.5 and e = 0.1, the minimum ratio of stellar
and planetary orbital semi-major axes is about four from
\citealt{Hol1999}), no planets have yet been observed in this regime.
This could be attributed to the selection effect of close binaries
being excluded from planet searches, as described above.  However,
this could also provide evidence for protoplanetary disk truncation by
a close stellar companion, preventing planet formation in systems with
separation ratios close to the limits permitted by numerical
simulations.

\subsection {Stellar Companions Might Influence Eccentricity of
Planetary Orbits}

Eccentricities of exoplanet orbits are significantly higher than those
of planets in our Solar System \citep{Mar2005b}.  \citet{Tak2005}
investigated whether the Kozai mechanism can explain this entirely,
and concluded that other effects are also at play.  We investigated
the potential impact of close stellar companions on the eccentricity
of planetary orbits, as these would have a greater gravitational
influence on the planet's orbit, and potentially reduce the period of
Kozai cycles.  Figure~\ref{eccrat} shows the eccentricity of the
planetary orbits as a function of the ratio of projected stellar
separations to the a $\sin{i}$ of planetary orbits, and does not
conclusively demonstrate any relationship.  However, even though three
data points do not provide conclusive evidence, it is interesting to
note that the systems with ratios under 100 have a minimum
eccentricity of 0.2, while larger ratio systems have lower
eccentricities.

We also looked at the relationship between period and eccentricity of
planetary orbits in systems with and without stellar companions.
Figure~\ref{eccper} shows the eccentricity of planetary orbits versus
the orbital period.  Planet orbits in systems with confirmed stellar
companions are represented by filled squares, orbits with candidate
stellar companions are represented by open squares, and orbits in
single star systems are denoted by open circles.  \citet{Udry2004} and
\citet{Egg2004} presented similar plots and concluded that all the
planets with a period P $\lesssim$ 40 days orbiting in multiple-star
systems have an eccentricity smaller than 0.05, whereas longer period
planets found in multiple-star systems can have larger eccentricities.
Our updated results show that this conclusion is no longer strictly
true.  The latest planet reported around 55 Cnc, designated with
suffix e, has a period of 2.81 days and an eccentricity of 0.17.
Also, we report HD 38529 as a multiple star system, which was assumed
to be a single star system in \citet{Udry2004}.  Planet HD 38529b has
a period of 14.31 days and an eccentricity of 0.29.  It appears that
single-star and multiple-star planetary systems have similar
period-eccentricity relationships.

\section {Conclusion}

Our comprehensive investigation of 131 exoplanet systems reveals that
30 (23\%) of these have stellar companions, an increase from 15
reported in previous such comprehensive efforts
\citep{Egg2004,Udry2004}.  We report new stellar companions to HD
38529 and HD 188015, and identify a candidate companion to HD 169830.
Our synthesis effort, bringing together disparate databases,
recognizes, for the first time, five additional stellar companions to
exoplanet hosts, including one triple system.  A by-product of our CPM
investigation is the determination that 20 of the WDS entries for
exoplanet hosts are not gravitationally bound to their ``primaries'',
but are chance alignments in the sky.  Some interesting examples in
the inventory of multiple-star exoplanet systems include: (1) At least
3 and possibly 5 exoplanet systems are stellar triples (see
\S\ref{sec:Discussion}); (2) Three systems (GJ 86, HD 41004, and
$\gamma$ Cep) have planets at $\sim$ Mercury to Mars distances and
potentially close-in stellar companions at projected separations
similar to the distance between the Sun and Uranus ($\sim$ 20 AU); (3)
Two systems (GJ 86 and HD 147513) have white dwarf companions.  These
results show that planets form and survive in a variety of stellar
multiplicity environments.  We hope that this compendium of stellar
multiples in exoplanet systems will provide a valuable benchmark for
future companion searches and exoplanet system analyses.

\section {Acknowledgments}

We wish to thank Charlie Finch and Jennifer Winters for their
supporting work in this effort.  We also thank G. Fritz Benedict and
Jacob Bean for obtaining and reducing the radial velocity data for HD
38529B.  We are grateful to Geoff Marcy and Kevin Apps for reviewing
our draft and providing useful suggestions, and to the anonymous
referee, who provided detailed comments based on a thorough review.
The photometric and spectroscopic observations reported here were
carried out under the auspices of the SMARTS (Small and Moderate
Aperture Research Telescope System) Consortium, which operates several
small telescopes at CTIO, including the 0.9m, 1.0m, and 1.5m.  TJH's
Space Interferometry Mission grant supported some of the work carried
out here.  This effort used multi-epoch images from the Digitized Sky
Survey, which was produced at the Space Telescope Science Institute
under U.S. Government grant NAG W-2166.  This work also used data
products from the Two Micron All Sky Survey, which is a joint project
of the University of Massachusetts and the Infrared Processing and
Analysis Center/California Institute of Technology, funded by the
National Aeronautics and Space Administration and the National Science
Foundation.  Additionally, this work used data from the SuperCOSMOS
Sky Survey, the Hipparcos catalog, and SIMBAD databases.

\clearpage


\begin{deluxetable}{lrrrrrlcc}
\tabletypesize{\footnotesize}
\tablecaption{Sample List of Exoplanet Systems Searched for Companions.
\label{Sample}}
\tablewidth{0pt}

\tablehead{
           \colhead{HD Name}&
           \multicolumn{2}{c}{Proper Motion}&
           \multicolumn{2}{c}{DSS Images}&
           \colhead{Total $\mu$}&
           \colhead{$\mu$ obs?}&
           \multicolumn{2}{c}{Companions} \\

           \colhead{}&
           \colhead{$\arcsec$ yr$^{-1}$}&
           \colhead{\sdeg}&
           \colhead{Epoch 1}&
           \colhead{Epoch 2}&
           \colhead{$\arcsec$}&
           \colhead{}&
           \colhead{CPM}&
           \colhead{Other} \\

           \colhead{(1)}&
           \colhead{(2)}&
           \colhead{(3)}&
           \colhead{(4)}&
           \colhead{(5)}&
           \colhead{(6)}&
           \colhead{(7)}&
           \colhead{(8)}&
           \colhead{(9)}}

\startdata

BD$-$10 3166 & 0.183 &  268.5 & 1983.29 & 1992.04 &  1.602 & YES &      &         \\
GJ 436       & 1.211 &  132.2 & 1955.28 & 1996.38 & 49.770 & YES &      &         \\
GJ 876       & 1.174 &  125.1 & 1983.76 & 1989.83 &  7.116 & YES &      &         \\
HD 000142    & 0.577 &   94.0 & 1982.87 & 1996.62 &  7.933 & YES &      & B       \\
HD 001237    & 0.438 &   97.6 & 1977.77 & 1997.58 &  8.676 & YES &      &         \\
HD 002039    & 0.080 &   79.0 & 1978.82 & 1997.61 &  1.503 & NO  &      &         \\
HD 002638    & 0.248 &  205.5 & 1983.53 & 1993.85 &  2.560 & YES &      &         \\
HD 003651    & 0.592 &  231.2 & 1953.91 & 1987.65 & 19.972 & YES &      &         \\
HD 004203    & 0.176 &  134.7 & 1954.00 & 1987.65 &  5.922 & YES &      &         \\
HD 004208    & 0.348 &   64.4 & 1980.63 & 1989.74 &  3.171 & NO  &      &         \\
HD 006434    & 0.554 &  197.8 & 1976.89 & 1990.73 &  7.666 & YES &      &         \\
HD 008574    & 0.298 &  122.1 & 1949.98 & 1991.76 & 12.453 & YES &      &         \\
HD 008673    & 0.250 &  109.8 & 1954.67 & 1991.76 &  9.273 & YES &      & B?      \\
HD 009826    & 0.418 &  204.4 & 1953.71 & 1989.77 & 15.073 & YES & B    & B       \\
HD 010647    & 0.198 &  122.6 & 1977.92 & 1997.61 &  3.898 & YES &      &         \\
HD 010697    & 0.115 &  203.1 & 1954.89 & 1986.69 &  3.657 & YES &      &         \\
HD 011964    & 0.441 &  236.6 & 1982.63 & 1991.70 &  4.003 & YES & B    & B       \\
HD 011977    & 0.105 &   46.1 & 1976.67 & 1987.72 &  1.160 & NO  &      &         \\
HD 012661    & 0.206 &  211.6 & 1953.87 & 1990.87 &  7.622 & NO  &      &         \\
HD 013189    & 0.006 &   13.3 & 1954.76 & 1989.83 &  0.210 & NO  &      &         \\
HD 013445    & 2.193 &   72.6 & 1975.85 & 1988.91 & 28.646 & YES &      & B       \\
HD 016141    & 0.464 &  199.7 & 1982.79 & 1997.74 &  6.937 & YES &      & B?      \\
HD 017051    & 0.399 &   56.7 & 1977.78 & 1997.81 &  7.995 & YES &      &         \\
HD 019994    & 0.205 &  109.7 & 1951.69 & 1997.84 &  9.463 & YES &      & B       \\
HD 020367    & 0.118 &  241.2 & 1953.77 & 1993.72 &  4.714 & YES &      &         \\
HD 022049    & 0.977 &  277.1 & 1982.79 & 1998.97 & 15.806 & YES &      &         \\
HD 023079    & 0.214 &  244.6 & 1978.82 & 1993.96 &  3.241 & YES &      &         \\
HD 023596    & 0.058 &   68.5 & 1953.03 & 1989.76 &  2.130 & NO  &      &         \\
HD 027442    & 0.175 &  196.0 & 1983.04 & 1997.74 &  2.573 & YES &      & B       \\
HD 027894    & 0.328 &   33.8 & 1983.04 & 1997.74 &  4.823 & YES &      &         \\
HD 028185    & 0.101 &  126.7 & 1982.82 & 1985.96 &  0.317 & NO  &      &         \\
HD 030177    & 0.067 &  100.3 & 1983.04 & 1997.74 &  0.985 & NO  &      &         \\
HD 033636    & 0.227 &  127.2 & 1954.85 & 1990.81 &  8.164 & YES &      &         \\
HD 034445    & 0.149 &  184.4 & 1954.85 & 1990.82 &  5.360 & YES &      &         \\
HD 037124    & 0.427 &  190.8 & 1951.91 & 1991.80 & 17.032 & YES &      &         \\
HD 037605    & 0.252 &  167.5 & 1955.90 & 1992.06 &  9.114 & YES &      &         \\
HD 038529    & 0.163 &  209.4 & 1951.91 & 1990.87 &  6.350 & YES & B    &         \\
HD 039091    & 1.096 &   16.5 & 1978.03 & 1989.99 & 13.116 & YES &      &         \\
HD 040979    & 0.179 &  148.0 & 1953.12 & 1989.83 &  6.570 & YES & B    & B       \\
HD 041004    & 0.078 &  327.0 & 1978.03 & 1993.96 &  1.243 & YES &      & B,C     \\
HD 045350    & 0.069 &  219.3 & 1953.19 & 1986.91 &  2.326 & NO  &      &         \\
HD 046375    & 0.150 &  130.3 & 1953.94 & 1998.88 &  6.740 & YES &      & B       \\
HD 047536    & 0.126 &   59.5 & 1979.00 & 1992.99 &  1.763 & MAR &      &         \\
HD 049674    & 0.128 &  164.1 & 1953.19 & 1989.86 &  4.694 & YES &      &         \\
HD 050499    & 0.097 &  314.8 & 1976.89 & 1994.21 &  1.679 & NO  &      &         \\
HD 050554    & 0.103 &  201.2 & 1956.27 & 1994.03 &  3.889 & YES &      &         \\
HD 052265    & 0.141 &  304.8 & 1983.04 & 1989.18 &  0.864 & NO  &      &         \\
HD 059686    & 0.087 &  150.5 & 1953.02 & 1989.08 &  3.137 & MAR &      &         \\
HD 063454    & 0.045 &  207.5 & 1975.94 & 1992.99 &  0.767 & NO  &      &         \\
HD 065216    & 0.190 &  320.1 & 1976.25 & 1991.13 &  2.827 & NO  &      &         \\
HD 068988    & 0.132 &   76.1 & 1954.01 & 1989.98 &  4.747 & YES &      &         \\
HD 070642    & 0.303 &  318.1 & 1976.97 & 1991.10 &  4.283 & NO  &      &         \\
HD 072659    & 0.150 &  229.2 & 1954.97 & 1992.03 &  5.559 & YES &      &         \\
HD 073256    & 0.192 &  290.0 & 1977.22 & 1991.26 &  2.697 & MAR &      &         \\
HD 073526    & 0.173 &  339.5 & 1977.06 & 1991.27 &  2.459 & MAR &      &         \\
HD 074156    & 0.202 &  17209 & 1953.02 & 1991.10 &  7.692 & YES &      &         \\
HD 075289    & 0.229 &  185.1 & 1977.06 & 1991.27 &  3.255 & YES &      & B       \\
HD 075732    & 0.539 &  244.2 & 1953.94 & 1998.30 & 23.908 & YES & B    & B       \\
HD 076700    & 0.308 &  293.2 & 1976.26 & 1991.05 &  4.558 & YES &      &         \\
HD 080606    & 0.047 &   81.6 & 1953.13 & 1995.25 &  1.979 & YES & B    & B       \\
HD 082943    & 0.174 &  179.2 & 1983.36 & 1987.32 &  0.689 & NO  &      &         \\
HD 083443    & 0.123 &  169.5 & 1980.06 & 1995.09 &  1.849 & NO  &      &         \\
HD 088133    & 0.264 &  182.8 & 1955.23 & 1998.99 & 11.555 & YES &      &         \\
HD 089307    & 0.276 &  261.8 & 1950.29 & 1987.32 & 10.219 & YES &      &         \\
HD 089744    & 0.183 &  220.9 & 1953.21 & 1990.23 &  6.773 & NO  &      & B       \\
HD 092788    & 0.223 &  183.2 & 1982.37 & 1991.21 &  1.971 & YES &      &         \\
HD 093083    & 0.177 &  211.6 & 1980.21 & 1995.10 &  2.636 & YES &      &         \\
HD 095128    & 0.321 &  279.9 & 1955.22 & 1998.38 & 13.855 & YES &      &         \\
HD 099492    & 0.755 &  284.7 & 1955.29 & 1996.28 & 30.944 & YES & A    & A       \\
HD 101930    & 0.348 &    2.5 & 1987.20 & 1992.24 &  1.754 & NO  &      &         \\
HD 102117    & 0.094 &  222.1 & 1987.20 & 1992.24 &  0.474 & NO  &      &         \\
HD 104985    & 0.174 &  122.1 & 1955.17 & 1997.11 &  7.299 & YES &      &         \\
HD 106252    & 0.280 &  175.1 & 1955.29 & 1991.27 & 10.076 & YES &      &         \\
HD 108147    & 0.192 &  251.5 & 1987.26 & 1996.30 &  1.735 & NO  &      &         \\
HD 108874    & 0.157 &  124.7 & 1955.39 & 1991.07 &  5.602 & YES &      &         \\
HD 111232    & 0.116 &   13.9 & 1987.08 & 1996.29 &  1.067 & NO  &      & B?      \\
HD 114386    & 0.353 &  203.0 & 1975.41 & 1992.25 &  5.943 & YES &      &         \\
HD 114729    & 0.369 &  213.2 & 1978.13 & 1991.21 &  4.826 & YES &      & B       \\
HD 114762    & 0.583 &  269.8 & 1950.30 & 1996.30 & 26.822 & YES &      & B       \\
HD 114783    & 0.138 &  274.0 & 1956.27 & 1996.23 &  5.514 & YES &      &         \\
HD 117176    & 0.622 &  202.2 & 1955.38 & 1997.35 & 26.110 & YES &      &         \\
HD 117207    & 0.217 &  250.7 & 1975.27 & 1991.21 &  3.458 & MAR &      &         \\
HD 117618    & 0.127 &  168.6 & 1975.19 & 1991.23 &  2.037 & NO  &      &         \\
HD 120136    & 0.483 &  276.4 & 1954.25 & 1992.20 & 18.328 & YES &      & B       \\
HD 121504    & 0.264 &  251.5 & 1987.26 & 1994.19 &  1.828 & NO  &      &         \\
HD 128311    & 0.323 &  140.5 & 1950.28 & 1989.25 & 12.588 & YES &      &         \\
HD 130322    & 0.191 &  222.6 & 1980.22 & 1996.37 &  3.085 & YES &      &         \\
HD 134987    & 0.400 &   86.1 & 1976.42 & 1991.50 &  6.034 & YES &      &         \\
HD 136118    & 0.126 &  280.7 & 1955.30 & 1992.41 &  4.676 & YES &      &         \\
HD 137759    & 0.019 &  334.5 & 1953.46 & 1995.15 &  0.792 & NO  &      &         \\
HD 141937    & 0.100 &   76.1 & 1976.41 & 1991.61 &  1.520 & NO  &      &         \\
HD 142022    & 0.339 &  264.7 & 1977.63 & 1996.30 &  6.329 & YES & B    & B       \\
HD 142415    & 0.153 &  228.1 & 1988.30 & 1992.58 &  0.654 & NO  &      &         \\
HD 143761 \tablenotemark{1} & 0.798 &  194.3 & 1950.28 & 1994.37 & 35.182 & YES &      &         \\
HD 145675    & 0.326 &  156.1 & 1955.23 & 1991.43 & 11.802 & YES &      &         \\
HD 147513    & 0.073 &   87.3 & 1987.39 & 1993.25 &  0.428 & NO  &      & B       \\
HD 149026    & 0.094 &  304.7 & 1954.49 & 1993.33 &  3.651 & YES &      &         \\
HD 150706    & 0.130 &  132.6 & 1955.39 & 1996.54 &  5.350 & YES &      & B?      \\
HD 154857    & 0.103 &  122.4 & 1987.30 & 1993.32 &  0.621 & NO  &      &         \\
HD 160691    & 0.192 &  184.5 & 1987.70 & 1992.58 &  0.938 & NO  &      &         \\
HD 162020    & 0.033 &  140.2 & 1987.71 & 1991.68 &  0.131 & NO  &      &         \\
HD 168443    & 0.242 &  202.3 & 1978.65 & 1988.59 &  2.406 & NO  &      &         \\
HD 168746    & 0.073 &  197.7 & 1978.65 & 1988.59 &  0.726 & NO  &      &         \\
HD 169830    & 0.015 &  356.8 & 1987.38 & 1992.41 &  0.075 & NO  &      & B?      \\
HD 177830    & 0.066 &  218.1 & 1950.46 & 1992.42 &  2.770 & NO  &      &         \\
HD 178911B   & 0.203 &   18.6 & 1955.39 & 1992.44 &  7.523 & YES & A    & A,C     \\
HD 179949    & 0.153 &  131.6 & 1987.42 & 1991.62 &  0.643 & NO  &      &         \\
HD 183263    & 0.038 &  208.2 & 1950.61 & 1992.59 &  1.595 & NO  &      &         \\
HD 186427    & 0.212 &  219.6 & 1951.53 & 1991.53 &  8.679 & YES & A    & A,C     \\
HD 187123    & 0.189 &  130.7 & 1952.54 & 1992.67 &  7.583 & YES &      &         \\
HD 188015    & 0.106 &  149.4 & 1953.53 & 1992.49 &  4.130 & YES & B    &         \\
HD 190228    & 0.126 &  123.7 & 1953.53 & 1992.49 &  4.910 & YES &      &         \\
HD 190360    & 0.861 &  127.5 & 1953.53 & 1992.49 & 33.549 & YES & B    & B       \\
HD 192263    & 0.270 &  346.4 & 1951.58 & 1988.67 & 10.013 & YES &      &         \\
HD 195019    & 0.354 &   99.2 & 1951.52 & 1990.71 & 13.874 & YES &      & B       \\
HD 196050    & 0.201 &  251.4 & 1977.61 & 1991.75 &  2.842 & MAR &      & B       \\
HD 196885    & 0.096 &   29.7 & 1953.68 & 1987.50 &  3.246 & YES &      &         \\
HD 202206    & 0.126 &  197.7 & 1977.55 & 1991.74 &  1.788 & NO  &      &         \\
HD 208487    & 0.156 &  139.3 & 1980.55 & 1995.63 &  2.353 & MAR &      &         \\
HD 209458    & 0.034 &  122.4 & 1950.54 & 1990.73 &  1.366 & NO  &      &         \\
HD 210277    & 0.458 &  169.2 & 1979.72 & 1987.79 &  3.693 & YES &      &         \\
HD 213240    & 0.236 &  214.9 & 1980.77 & 1995.65 &  3.510 & YES & B    & B       \\
HD 216435    & 0.232 &  110.6 & 1980.54 & 1996.62 &  3.730 & NO  &      &         \\
HD 216437    & 0.085 &  329.5 & 1978.82 & 1996.79 &  1.527 & NO  &      &         \\
HD 216770    & 0.290 &  127.9 & 1980.78 & 1995.79 &  4.354 & YES &      &         \\
HD 217014    & 0.217 &   73.7 & 1954.59 & 1990.79 &  7.856 & YES &      &         \\
HD 217107    & 0.017 &  200.7 & 1982.80 & 1991.68 &  0.151 & NO  &      & B?      \\
HD 219449    & 0.369 &   92.6 & 1983.82 & 1991.76 &  2.931 & YES & B    & B,C     \\
HD 222404    & 0.136 &  339.0 & 1954.73 & 1992.76 &  5.172 & YES &      & B       \\
HD 222582    & 0.183 &  232.6 & 1983.54 & 1989.83 &  1.152 & YES & B    & B       \\
HD 330075    & 0.254 &  248.2 & 1988.45 & 1995.25 &  1.725 & NO  &      &         \\

\enddata

\tablenotetext{1}{We conclude that this system ($\rho$ CrB) has either
a planetary or a stellar companion, but not both.  See
\S\ref{sec:Hip} for more details.}

\end{deluxetable}

\clearpage


\begin{deluxetable}{rllccrlrrrrrrrcll}
\rotate 
\setlength{\tabcolsep}{0.05in}
\tabletypesize{\scriptsize} 
\tablecaption{Exoplanet Systems with Stellar Companions.
\label{results}}
\tablewidth{0pt}

\tablehead{\vspace{-25pt} \\
           \colhead{S}&
           \colhead{HD Name}&
           \colhead{Other Name}&
           \colhead{C}&
           \colhead{RA \hskip25pt DEC}&
           \colhead{$\pi$}&
           \colhead{Dist}&
           \colhead{}&
           \colhead{SpT}&
           \colhead{Ang Sep}&
           \colhead{PA}&
           \colhead{Proj. Sep.}&
           \colhead{$M \sin{i}$}&
           \colhead{a $\sin{i}$}&
           \colhead{e}&
           \colhead{Sources}&
           \colhead{References} \\

           \colhead{}&
           \colhead{}&
           \colhead{}&
           \colhead{}&
           \colhead{(J2000)}&
           \colhead{($\arcsec$)}&
           \colhead{(pc)}&
           \colhead{}&
           \colhead{}&
           \colhead{($\arcsec$)}&
           \colhead{(\sdeg)}&
           \colhead{(AU)}&
           \colhead{(\mjup)}&
           \colhead{(AU)}&
           \colhead{}&
           \colhead{}&
           \colhead{} \\

           \colhead{(1)}&
           \colhead{(2)}&
           \colhead{(3)}&
           \colhead{(4)}&
           \colhead{(5)}&
           \colhead{(6)}&
           \colhead{(7)}&
           \colhead{(8)}&
           \colhead{(9)}&
           \colhead{(10)}&
           \colhead{(11)}&
           \colhead{(12)}&
           \colhead{(13)}&
           \colhead{(14)}&
           \colhead{(15)}&
           \colhead{(16)}&
           \colhead{(17)}}
\startdata

1  & 000142 & GJ 4.2         & A  & 00 06 19.18 $-$49 04 30.7 & 0.03900 & 25.6 & T & G1IV   &      &     &       &       &       &       &      &         \\
   &        &                & b  &                           &         &      &   &        &      &     &       & 1     & 0.98  & 0.38  &      &         \\
   &        &                & B  &                           &         &      &   &        & 5.4  & 177 & 138   &       &       &       & WC   & 1,2     \\
2  & 009826 & $\upsilon$ And & A  & 01 36 47.84 +41 24 19.7   & 0.07425 & 13.5 & T & F8.0V  &      &     &       &       &       &       &      &         \\
   &        &                & b  &                           &         &      &   &        &      &     &       & 0.69  & 0.059 & 0.012 &      &         \\
   &        &                & c  &                           &         &      &   &        &      &     &       & 1.89  & 0.829 & 0.28  &      &         \\
   &        &                & d  &                           &         &      &   &        &      &     &       & 3.75  & 2.53  & 0.27  &      &         \\
   &        &                & B  & 01 36 50.40 +41 23 32.1   &         &      &   & M4.5V  & 52   & 150 & 702   &       &       &       & P    & 2,3,4   \\
3  & 011964 & GJ 81.1        & A  & 01 57 09.61 $-$10 14 32.7 & 0.02943 & 34.0 & T & G5     &      &     &       &       &       &       &      &         \\
   &        &                & b  &                           &         &      &   &        &      &     &       & 0.11  & 0.229 & 0.15  &      &         \\
   &        &                & c  &                           &         &      &   &        &      &     &       & 0.7   & 3.167 & 0.3   &      &         \\
   &        &                & B \tablenotemark{1} & 01 57 11.07 $-$10 14 53.2 &   &  &   & & 29.7 & 133 & 1010  &       &       &       & PWC  & 5,6     \\
4  & 013445 & GJ 86          & A  & 02 10 25.93 $-$50 49 25.4 & 0.09163 & 10.9 & T & K1V    &      &     &       &       &       &       &      &         \\
   &        &                & b  &                           &         &      &   &        &      &     &       & 4.01  & 0.11  & 0.046 &      &         \\ 
   &        &                & B  &                           &         &      &   & wd     & 1.93 & 119 & 21    &       &       &       & O    & 7,8,9   \\
5  & 019994 & GJ 128         & A  & 03 12 46.44 $-$01 11 46.0 & 0.04469 & 22.4 & T & F8.5V  &      &     &       &       &       &       &      &         \\
   &        &                & b  &                           &         &      &   &        &      &     &       & 2     & 1.3   & 0.2   &      &         \\
   &        &                & B  &                           &         &      &   & M      & 2.5  & 213 & 56    &       &       &       & WCD  & 10,11,12,13 \\
6  & 027442 & $\epsilon$ Ret & A  & 04 16 29.03 $-$59 18 07.8 & 0.05484 & 18.2 & T & K2IVa  &      &     &       &       &       &       &      &         \\
   &        &                & b  &                           &         &      &   &        &      &     &       & 1.28  & 1.18  & 0.07  &      &         \\
   &        &                & B \tablenotemark{1} &          &         &      &   &        & 13.8 & 36  & 251   &       &       &       & WCI  & 14,15   \\
7  & 038529 & HIP 27253      & A  & 05 46 34.91 +01 10 05.5   & 0.02357 & 42.4 & T & G4V    &      &     &       &       &       &       &      & 16      \\
   &        &                & b  &                           &         &      &   &        &      &     &       & 0.78  & 0.129 & 0.29  &      &         \\
   &        &                & c  &                           &         &      &   &        &      &     &       & 12.7  & 3.68  & 0.36  &      &         \\
   &        &                & B \tablenotemark{2}  & 05 46 19.38 +01 12 47.2   &         & 28.7 & C & M3.0V  & 284  & 305 & 12042 &       &       &       & P    &         \\
8  & 041004 & HIP 28393      & A  & 05 59 49.65 $-$48 14 22.9 & 0.02324 & 43.0 & T & K1V    &      &     &       &       &       &       &      &         \\
   &        &                & b  &                           &         &      &   &        &      &     &       & 2.3   & 1.31  & 0.39  &      &         \\
   &        &                & B  & 05 59 43.81 $-$48 12 11.9 &         &      &   & M2.5V  & 0.5  & 176 & 22    &       &       &       & WH   & 4,17,18,19 \\
   &        &                & C  &                           &         &      &   &        &      &     &       & 18.4  & 0.016 & 0.08  &      & 18,19    \\
9  & 040979 &  BD+44 1353    & A  & 06 04 29.95 +44 15 37.6   & 0.03000 & 33.3 & T & F8     &      &     &       &       &       &       &      &         \\
   &        &                & b  &                           &         &      &   &        &      &     &       & 3.32  & 0.811 & 0.23  &      &         \\
   &        &  BD+44 1351    & B  & 06 04 13.02 +44 16 41.1   &         & 15.2 & P & K5     & 192  & 290 & 6394  &       &       &       & P    & 4,16,20,21 \\
10 & 046375 & HIP 31246      & A  & 06 33 12.62 +05 27 46.5   & 0.02993 & 33.4 & T & K0V    &      &     &       &       &       &       &      &         \\
   &        &                & b  &                           &         &      &   &        &      &     &       & 0.249 & 0.041 & 0.04  &      &         \\
   &        &                & B \tablenotemark{1} & 06 33 12.10 +05 27 53.2 & & 26.4 & C & & 9.4  & 308 & 314   &       &       &       & WI   & 22,23   \\
11 & 075289 & HIP 43177      & A  & 08 47 40.39 -41 44 12.5   & 0.03455 & 28.9 & T & G0V    &      &     &       &       &       &       &      &         \\
   &        &                & b  &                           &         &      &   &        &      &     &       & 0.42  & 0.046 & 0.054 &      &         \\
   &        &                & B  & 08 47 42.26 -41 44 07.6   &         &      &   &        & 21.5 & 78  & 621   &       &       &       & O    & 24      \\
12 & 075732 & 55 Cnc         & A  & 08 52 35.81 +28 19 50.9   & 0.07980 & 12.5 & T & K0IV-V &      &     &       &       &       &       &      &         \\
   &        &                & e  &                           &         &      &   &        &      &     &       & 0.045 & 0.038 & 0.174 &      &         \\
   &        &                & b  &                           &         &      &   &        &      &     &       & 0.784 & 0.115 & 0.020 &      &         \\
   &        &                & c  &                           &         &      &   &        &      &     &       & 0.217 & 0.24  & 0.44  &      &         \\
   &        &                & d  &                           &         &      &   &        &      &     &       & 3.92  & 5.257 & 0.327 &      &         \\
   &        &                & B  & 08 52 40.85 +28 18 59.0   &         &  8.7 & C & M4     & 84   & 130 & 1050  &       &       &       & PWCD & 4,12,25,26,27 \\
13 & 080606 & HIP 45982      & A  & 09 22 37.57 +50 36 13.4   & 0.01713 & 58.4 & T & G5     &      &     &       &       &       &       &      &         \\
   &        &                & b  &                           &         &      &   &        &      &     &       & 3.41  & 0.439 & 0.927 &      &         \\
   & 080607 & HIP 45983      & B  & 09 22 39.73 +50 36 13.9   &         &      &   & G5     & 20.6 & 269 & 1203  &       &       &       & PWH  & 4,28    \\
14 & 089744 & HIP 50786      & A  & 10 22 10.56 +41 13 46.3   & 0.02565 & 39.0 & T & F8IV   &      &     &       &       &       &       &      &         \\
   &        &                & b  &                           &         &      &   &        &      &     &       & 7.99  & 0.89  & 0.67  &      &         \\
   &        &                & B  & 10 22 14.87 +41 14 26.4   &         &      &   & L0V    & 63.0 & 48  & 2456  &       &       &       & O    & 29,30   \\
15 & 099492 & GJ 429B        & B  & 11 26 46.28 +03 00 22.8   & 0.05559 & 18.0 & T & K2V    &      &     &       &       &       &       &      &         \\
   &        &                & b  &                           &         &      &   &        &      &     &       & 0.122 & 0.119 & 0.05  &      &         \\
   & 099491 & GJ 429A        & A  & 11 26 45.32 +03 00 47.2   & 0.05659 & 17.7 & T & K0IV   & 28.6 & 150 & 515   &       &       &       & PWHC & 31      \\
16 & 114729 & HIP 64459      & A  & 13 12 44.26 $-$31 52 24.1 & 0.02857 & 35.0 & T & G3V    &      &     &       &       &       &       &      &         \\
   &        &                & b  &                           &         &      &   &        &      &     &       & 0.82  & 2.08  & 0.31  &      &         \\
   &        &                & B  & 13 12 43.97 $-$31 52 17.0 &         &      &   &        & 8.05 & 333 & 282   &       &       &       & O    & 32      \\
17 & 114762 & HIP 64426      & A  & 13 12 19.74 +17 31 01.6   & 0.02465 & 40.6 & T & F9V    &      &     &       &       &       &       &      &         \\
   &        &                & b  &                           &         &      &   &        &      &     &       & 11.02 & 0.3   & 0.25  &      &         \\
   &        &                & B  &                           &         &      &   &        & 3.26 & 30  & 132   &       &       &       & O    & 3,4     \\
18 & 120136 & $\tau$ Boo     & A  & 13 47 15.74 +17 27 24.9   & 0.06412 & 15.6 & T & F6IV   &      &     &       &       &       &       &      &         \\
   &        &                & b  &                           &         &      &   &        &      &     &       & 4.13  & 0.05  & 0.01  &      &         \\
   &        &                & B  &                           &         &      &   &        & 2.87 & 31  & 45    &       &       &       & WCD  & 3,4,12  \\
19 & 142022 & GJ 606.1       & A  & 16 10 15.02 $-$84 13 53.8 & 0.02788 & 38.9 & T & G8/K0V &      &     &       &       &       &       &      &         \\
   &        &                & b  &                           &         &      &   &        &      &     &       & 4.4   & 2.8   & 0.57  &      &         \\
   &        &                & B  & 16 10 25.34 $-$84 14 06.7 &         &      &   & K7V    & 20.4 & 130 & 794   &       &       &       & PWC  & 33,34   \\
20 & 147513 & GJ 620.1       & A  & 16 24 01.29 $-$39 11 34.7 & 0.07769 & 12.9 & T & G5V    &      &     &       &       &       &       &      &         \\
   &        &                & b  &                           &         &      &   &        &      &     &       & 1     & 1.26  & 0.52  &      &         \\
   &        &                & B  & 16 23 33.83 $-$39 13 46.1 & 0.07804 & 12.8 & T & wd     & 345  & 245 & 4451  &       &       &       & C    & 13,35   \\
21 & 178911B & HIP 94076B     & B  & 19 09 03.10 +34 35 59.5   & 0.02140 & 46.7 & T & G5     &      &     &       &       &       &       &      &         \\
   &        &                & b  &                           &         &      &   &        &      &     &       & 6.292 & 0.32  & 0.124 &      &         \\
   & 178911 & HIP 94076      & A  & 19 09 04.38 +34 36 01.6   & 0.02042 & 49.0 & T & G1V J  & 16.1 &  82 & 789   &       &       &       & PWH  & 4,36,37,38,39 \\
   &        &                & C \tablenotemark{3} &                           &         &      &   &        & 0.1  & 21  & 4.9   &       &       &       & W    &         \\
22 & 186427 & 16 Cyg B       & B  & 19 41 51.97 +50 31 03.1   & 0.04670 &      &   & G3V    &      &     &       &       &       &       &      &         \\
   &        &                & b  &                           &         &      &   &        &      &     &       & 1.69  & 1.67  & 0.67  &      &         \\
   & 186408 & 16 Cyg A       & A  & 19 41 48.95 +50 31 30.2   & 0.04625 & 21.6 & T & G1.5V J & 39.8 & 313 & 860   &       &       &       & PWC  & 2,3,4,40,41 \\
   &        &                & C \tablenotemark{3} &                           &         &      &   &        & 3.4  & 209 & 73    &       &       &       & W    &         \\
23 & 188015 & HIP 97769      & A  & 19 52 04.54 +28 06 01.4   & 0.01900 & 52.6 & T & G5IV   &      &     &       &       &       &       &      &         \\
   &        &                & b  &                           &         &      &   &        &      &     &       & 1.26  & 1.19  & 0.15  &      &         \\
   &        &                & B \tablenotemark{2} & 19 52 05.51 +28 06 03.7   &         & 46.9 & C &        & 13   &  85 & 684   &       &       &       & P    &         \\
24 & 190360 & GJ 777         & A  & 20 03 37.41 +29 53 48.5   & 0.06292 & 15.9 & T & G7IV-V &      &     &       &       &       &       &      &         \\
   &        &                & c  &                           &         &      &   &        &      &     &       & 0.057 & 0.128 & 0.01  &      &         \\
   &        &                & b  &                           &         &      &   &        &      &     &       & 1.502 & 3.92  & 0.36  &      &         \\
   &        &                & B  & 20 03 26.58 +29 51 59.5   &         & 18.5 & P & M4.5V  & 179  & 234 & 2846  &       &       &       & PWC  & 4,5,16,42 \\
25 & 195019 & HIP 100970     & A  & 20 28 18.64 +18 46 10.2   & 0.02677 & 37.3 & T & G3IV-V &      &     &       &       &       &       &      &         \\
   &        &                & b  &                           &         &      &   &        &      &     &       & 3.43  & 0.14  & 0.05  &      &         \\
   &        &                & B  &                           &         &      &   &        & 3.5  & 330 & 131   &       &       &       & W    & 4,5,43,44 \\
26 & 196050 & HIP 101806     & A  & 20 37 51.71 $-$60 38 04.1 & 0.02131 & 46.9 & T & G3V    &      &     &       &       &       &       &      &         \\
   &        &                & b  &                           &         &      &   &        &      &     &       & 3     & 2.5   & 0.28  &      &         \\
   &        &                & B  & 20 37 51.85 $-$60 38 14.9 &         &      &   &        & 10.9 & 175 & 510   &       &       &       & O    & 32      \\
27 & 213240 & HIP 111143     & A  & 22 31 00.37 $-$49 25 59.8 & 0.02454 & 40.8 & T & G0/G1V &      &     &       &       &       &       &      &         \\
   &        &                & b  &                           &         &      &   &        &      &     &       & 4.5   & 2.03  & 0.45  &      &         \\
   &        &                & B  & 22 31 08.26 $-$49 26 56.7 &         & 41.8 & C & M5.0V  & 95.8 & 127 & 3909  &       &       &       & P    & 32      \\
28 & 219449 & GJ 893.2       & A  & 23 15 53.49 $-$09 05 15.9 & 0.02197 & 45.5 & T & K0III  &      &     &       &       &       &       &      &         \\
   &        &                & b  &                           &         &      &   &        &      &     &       & 2.9   & 0.3   & $-$   &      &         \\
   & 219430 &                & B  \tablenotemark{1} & 23 15 51.00 $-$09 04 42.7 &         & 42.4 \tablenotemark{4} & C & K8V J   & 49.4 & 313 & 2248  &       &       &       & PWC  & 6,45    \\
   &        &                & C  \tablenotemark{1,5} &                         &         & 42.4 \tablenotemark{4} & C &         & 0.4  & 101 &   18  &       &       &       & W    &         \\
29 & 222404 & $\gamma$ Cephei & A & 23 39 20.85 +77 37 56.2   & 0.07250 & 13.8 & T & K1III  &      &     &       &       &       &       &      &         \\
   &        &                & b  &                           &         &      &   &        &      &     &       & 1.59  & 2.03  & 0.2   &      &         \\
   &        &                & B  &                           &         &      &   &        & $-$  &     &       &       & 20.3  & 0.39  & H    & 4,46,47,48,49 \\
30 & 222582 & HIP 116906     & A  & 23 41 51.53 $-$05 59 08.7 & 0.02384 & 42.0 & T & G5     &      &     &       &       &       &       &      &         \\
   &        &                & b  &                           &         &      &   &        &      &     &       & 5.11  & 1.35  & 0.76  &      &         \\
   &        &                & B \tablenotemark{1} & 23 41 45.14 $-$05 58 14.8 &         & 32.1 & C & M3.5V  & 113  & 302 & 4746  &       &       &       & PW   & 6       \\

\tableline 
\multicolumn{16}{c}{Candidate (Unconfirmed) Stellar Companions} \\
\tableline 

31 & 008673 & HIP 6702       & A  & 01 26 08.78 +34 34 46.9   & 0.02614 & 38.3 & T & F7V    &      &     &       &       &       &       &      &         \\
   &        &                & b  &                           &         &      &   &        &      &     &       & 14    & 1.58  & $-$   &      &         \\
   &        &                & B  &                           &         &      &   &        & 0.1  & 78  & 3.8   &       &       &       & W    &         \\
32 & 016141 & HIP 12048      & A  & 02 35 19.93 $-$03 33 38.2 & 0.02785 & 35.9 & T & G5IV   &      &     &       &       &       &       &      &         \\
   &        &                & b  &                           &         &      &   &        &      &     &       & 0.23  & 0.35  & 0.21  &      &         \\
   &        &                & B  & 02 35 19.88 $-$03 33 43.9 &         &      &   &        & 6.2  & 188 & 222   &       &       &       & O    & 32      \\
33 & 111232 & HIP 62534      & A  & 12 48 51.75 $-$68 25 30.5 & 0.03463 & 28.9 & T & G8V    &      &     &       &       &       &       &      &         \\
   &        &                & b  &                           &         &      &   &        &      &     &       & 6.8   & 1.97  & 0.2   &      &         \\
   &        &                & B  &                           &         &      &   &        &      &     &       &       &       &       & H    & 13      \\
34 & 150706 & GJ 632         & A  & 16 31 17.59 +79 47 23.2   & 0.03673 & 27.2 & T & G0     &      &     &       &       &       &       &      &         \\
   &        &                & b  &                           &         &      &   &        &      &     &       & 1     & 0.82  & 0.38  &      &         \\
   &        &                & B  &                           &         &      &   &        &      &     &       &       &       &       & H    & 50      \\
35 & 169830 & HIP 90485      & A  & 18 27 49.48 $-$29 49 00.7 & 0.02753 & 36.3 & T & F9V    &      &     &       &       &       &       &      &         \\
   &        &                & b  &                           &         &      &   &        &      &     &       & 2.88  & 0.81  & 0.31  &      &         \\
   &        &                & c  &                           &         &      &   &        &      &     &       & 4.04  & 3.6   & 0.33  &      &         \\
   &        &                & B  \tablenotemark{6} & 18 27 48.65 $-$29 49 01.6  &         &      &   &        & 11   & 270 & 399   &       &       &       &      &         \\
36 & 217107 & HIP 113421     & A  & 22 58 15.54 $-$02 23 43.4 & 0.05071 & 19.7 & T & G8IV-V &      &     &       &       &       &       &      &         \\
   &        &                & b  &                           &         &      &   &        &      &     &       & 1.37  & 0.074 & 0.13  &      &         \\
   &        &                & c  &                           &         &      &   &        &      &     &       & 2.1   & 4.3   & 0.55  &      &         \\
   &        &                & B  &                           &         &      &   &        & 0.3  & 156 & 6     &       &       &       & W    & 51,52   \\

\enddata

\tablenotetext{1}{Known companion, but first identification of the
star as a companion to an exoplanet host.}
\tablenotetext{2}{New stellar companion reported by this work.}
\tablenotetext{3}{Separation and position angle are listed with
respect to component A.  A and C have been referred to as Aa and Ab,
respectively in other publications, but we follow a consistent naming
convention, using uppercase letters to represent stars and lowercase
letters to denote planets.}
\tablenotetext{4}{Photometry obtained is for the BC pair.  Distance
estimate assumes identical binary components.}
\tablenotetext{5}{Separation and position angle are listed with
respect to component B.}
\tablenotetext{6}{New candidate companion reported by this work,
via Kevin Apps.}

\tablecomments{Planet data is from Exoplanet Encyclopedia web site
$http://vo.obspm.fr/exoplanetes/encyclo/catalog.php$.}
\tablerefs{
(1) \citet{Bai1900}; 
(2) \citet{Low2002}; 
(3) \citet{Pat2002}; 
(4) \citet{Egg2004}; 
(5) \citet{All2000}; 
(6) \citet{UCAC2};
(7) \citet{Els2001}; 
(8) \citet{Mug2005a};
(9) \citet{Que2000}; 
(10) \citet{Smy1844};
(11) \citet{Hale1994};
(12) \citet{Duq1991}; 
(13) \citet{May2004}; 
(14) \citet{Jes1955};
(15) \citet{Hold1966};
(16) \citet{Lep2005};
(17) \citet{See1896};
(18) \citet{Zuc2003};
(19) \citet{Zuc2004};
(20) \citet{Hog1998}; 
(21) \citet{Hal1986}; 
(22) \citet{Sou1985};
(23) \citet{Urb1998};
(24) \citet{Mug2004b};
(25) \citet{vanAlt1995};
(26) \citet{Dahn1988}; 
(27) \citet{Mar2002};
(28) \citet{Naef2001};
(29) \citet{Wil2001};
(30) \citet{Mug2004a};
(31) \citet{Mar2005a};
(32) \citet{Mug2005b};
(33) \citet{Luy1979};
(34) \citet{Egg2005};
(35) \citet{Weg1973};
(36) \citet{McAl1987a};
(37) \citet{Bal2004};
(38) \citet{Hart2000};
(39) \citet{Zuc2002}; 
(40) \citet{Turn2001};
(41) \citet{Coc1997};
(42) \citet{Naef2003}; 
(43) \citet{Hou1887};
(44) \citet{Fis1999};
(45) \citet{Wil1953};
(46) \citet{Msn2001}; 
(47) \citet{Cam1988}; 
(48) \citet{Gri2002}; 
(49) \citet{Hat2003};
(50) \citet{Hal2003}; 
(51) \citet{McAl1987b};
(52) \citet{Msn1999}.
}

\end{deluxetable}
\clearpage


\begin{deluxetable}{crcrrcrc}
\tabletypesize{\footnotesize}
\tablecaption{WDS Entries that are not Gravitationally Bound Companions. 
\label{WDS-no}}
\tablewidth{0pt}

\tablehead{
           \colhead{WDS ID}&
           \colhead{HD Name}&
           \colhead{Comp}&
           \colhead{$\theta$}&
           \colhead{$\rho$}&
           \colhead{Epoch}&
           \colhead{\#} &
           \colhead{Notes} \\

           \colhead{}&
           \colhead{}&
           \colhead{}&
           \colhead{\sdeg}&
           \colhead{$\arcsec$}&
           \colhead{}&
           \colhead{}&
           \colhead{} \\

           \colhead{(1)}&
           \colhead{(2)}&
           \colhead{(3)}&
           \colhead{(4)}&
           \colhead{(5)}&
           \colhead{(6)}&
           \colhead{(7)}&
           \colhead{(8)}}

\startdata

00394+2115   & 003651 &      &  80 & 167.6 & 1997 &  9 & 1 \\
01368+4124   & 009826 & AB   & 128 & 114.0 & 1909 &  1 & 1 \\
01368+4124   & 009826 & AC   & 289 & 273.6 & 1991 &  7 & 1 \\
03329$-$0927 & 022049 &      & 143 &   0.0 & 1975 &  1 & 2 \\
11268+0301   & 099492 & AC   & 187 &  90.5 & 1937 &  3 & 1 \\
13284+1347   & 117176 & AB   & 127 & 268.6 & 2002 & 13 & 1 \\
13284+1347   & 117176 & AC   & 263 & 325.5 & 1923 &  1 & 1 \\
13573$-$5602 & 121504 &      &  55 &  36.2 & 1999 & 32 & 3 \\
15249+5858   & 137759 &      &  50 & 254.8 & 2002 & 12 & 4 \\
16010+3318   & 143761 &      &  49 & 135.3 & 2002 & 22 & 1 \\
19091+3436   & 178911 & Aa-C & 130 &  60.0 & 1944 &  1 & 1 \\
20140$-$0052 & 192263 & A-BC & 102 &  73.1 & 2003 & 19 & 1 \\
20140$-$0052 & 192263 & AD   & 244 &  71.3 & 1921 &  1 & 1 \\
20140$-$0052 & 192263 & BC-D &  65 &  23.5 & 1998 &  8 & 1 \\
20283+1846   & 195019 & AC   &  72 &  70.9 & 1998 & 11 & 1 \\
20283+1846   & 195019 & AD   &  97 &  84.5 & 1998 &  2 & 1 \\
20399+1115   & 196885 &      &   6 & 182.9 & 2000 & 13 & 1 \\
22310$-$4926 & 213240 &      & 359 &  21.9 & 1999 &  7 & 1 \\
23159$-$0905 & 219449 & AD   & 274 &  80.4 & 1924 &  6 & 1 \\
23159$-$0905 & 219449 & BC-E & 341 &  19.7 & 1924 &  6 & 1 \\

\enddata

\tablecomments{Columns 1, 3 and 7 are listed here exactly as in WDS
catalog.  Columns 4, 5 and 6 correspond to the most recent
observation.  All data are as of June 20, 2005.  Certain pairs of
multiple systems omitted from this table are confirmed to be
gravitationally bound companions (01368+4124AD, 11268+0301AB,
19091+3436Aa \& Aa-B, 20283+1846AB, and 23159$-$0905A-BC \& BC).  One
omitted pair (20140$-$0052BC) has several speckle observations
\citep{Jon1911,Jon1917,Jon1944,Van1911,VanB1960}, and several failed
attempts \citep{vandb1949,vandb1960,vandb1963,Cou1953,Bai1957}, and is
hence inconclusive.
Column 8 notes:
(1) {DSS multi-epoch plates do not show CPM for WDS entry.  In fact,
  proper motion of the primary star causes change in separation and
  position angle indicating that the ``companion'' is a background
  star.}
(2) {Primary star is eps Eri, the well studied exoplanet system.  WDS
  listing is based on a single speckle measure by \citet{Bla1977}.
  This system has been observed 13 other times and no companion was
  resolved \citep{McAl1978,Hart1984,Opp2001}.}
(3) {Primary's $\mu$ = 0\farcs264 yr$^{-1}$ at 251\sdeg~from Hipparcos
  is not detectable in DSS plates.  For the WDS ``companion'',
  SuperCOSMOS lists $\mu$ = 0\farcs013 yr$^{-1}$ at 91\sdeg, clearly
  not matching the primary's.}
(4) {Primary does not show detectable proper motion in DSS plates.
  Planet discovery paper, \citet{Frink2002}, refuted the WDS entry
  based on distance estimate to WDS entry and proper motion
  comparisons.}
}

\end{deluxetable}
\clearpage
\thispagestyle{empty}
\setlength{\voffset}{15mm}
\begin{deluxetable}{llrrrrrrrrrrrrrrrr}
\rotate
\tabletypesize{\footnotesize}
\tablecaption{Observations and Computed Distances.
\label{Obs}}
\tablewidth{0pt}
\tablehead{
           \colhead{HD Name}&
           \colhead{SpT}&
           \multicolumn{3}{c}{Plate Mags.}&
           \multicolumn{3}{c}{CCD Mags.}&
           \colhead{\#}&
           \multicolumn{3}{c}{Infrared Mags.}&
           \colhead{D$_{plt}$}&
           \colhead{Err}&
           \colhead{\# Rel}&
           \colhead{D$_{CCD}$}&
           \colhead{Err}&
           \colhead{\# Rel} \\

           \colhead{}&
           \colhead{}&
           \colhead{$B$}&
           \colhead{$R$}&
           \colhead{$I$}&
           \colhead{$V$}&
           \colhead{$R$}&
           \colhead{$I$}&
           \colhead{}&
           \colhead{$J$}&
           \colhead{$H$}&
           \colhead{$K_S$}&
           \colhead{pc}&
           \colhead{pc}&
           \colhead{}&
           \colhead{pc}&
           \colhead{pc}&
           \colhead{} \\
           \colhead{(1)}&
           \colhead{(2)}&
           \colhead{(3)}&
           \colhead{(4)}&
           \colhead{(5)}&
           \colhead{(6)}&
           \colhead{(7)}&
           \colhead{(8)}&
           \colhead{(9)}&
           \colhead{(10)}&
           \colhead{(11)}&
           \colhead{(12)}&
           \colhead{(13)}&
           \colhead{(14)}&
           \colhead{(15)}&
           \colhead{(16)}&
           \colhead{(17)}&
           \colhead{(18)}}
\startdata
\tableline \vspace{-15pt} \\
\multicolumn{16}{c}{Exoplanet Host Without Parallax} \\
\tableline \vspace{-15pt} \\
BD$-$10 3166     &          &  9.90 &  8.80 &  8.08 & 10.03 &  9.59 &  9.19 & 1 &  8.61 &  8.30 &  8.12 &  33.8 &   8.8 &  1 &  66.8 &  10.0 &  1 \\
\tableline \vspace{-15pt} \\
\multicolumn{16}{c}{Confirmed Companions} \\
\tableline \vspace{-15pt} \\
HD 038529B  & M3.0V   & 13.81 & 11.84 & 10.05 & 13.35 & 12.29 & 10.98 & 3 &  9.72 &  9.04 &  8.80 &  31.8 &   9.0 & 11 &  28.7 &   4.8 & 12 \\
HD 040979B  &         &  9.92 &  8.72 &       &       &       &       &   &  7.27 &  6.79 &  6.69 &  15.2 &   4.0 &  3 &       &       &    \\
HD 046375B  &         &       &       &       & 11.80 & 11.01 &  9.80 & 3 &  8.70 &  8.08 &  7.84 &       &       &    &  26.4 &   6.0 & 12 \\
HD 075732B  &         & 13.14 & 11.53 &       & 13.26 & 11.91 & 10.24 & 2 &  8.56 &  7.93 &  7.67 &  14.5 &   4.6 &  6 &   8.7 &   1.4 & 12 \\
HD 188015B  &         &       &       &       &       & 15.54 & 13.91 & 1 & 12.09 & 11.59 & 11.34 &       &       &    &  46.9 &   9.5 &  7 \\
HD 190360B  &         & 15.30 & 12.35 &       &       &       &       &   &  9.55 &  9.03 &  8.71 &  18.5 &   6.2 &  6 &       &       &    \\
HD 213240B  & M5.0V   &       &       &       & 17.40 & 15.96 & 14.13 & 1 & 12.36 & 11.74 & 11.47 &       &       &    &  41.8 &   6.5 & 12 \\
HD 219449BC & Early K &       &       &       &  9.17 &  8.57 &  8.05 & 1 &  7.31 &  6.84 &  6.69 &       &       &    &  29.9 &   4.7 &  6 \\
HD 222582B  & M3.5V   & 15.25 & 13.16 & 11.41 & 14.49 & 13.33 & 11.83 & 1 & 10.39 &  9.81 &  9.58 &  35.1 &   9.3 & 11 &  32.1 &   5.0 & 12 \\
\tableline \vspace{-15pt} \\
\multicolumn{16}{c}{Candidate Companions} \\
\tableline \vspace{-15pt} \\
HD 169830B  &         &       &       &       & 14.35 & 13.62 & 12.39 & 1 & 10.16 &  9.50 &  9.35 &       &       &    &  29.2 &  23.4 & 12 \\
\tableline \vspace{-15pt} \\
\multicolumn{16}{c}{Refuted Candidate Companions} \\
\tableline \vspace{-15pt} \\
BD$-$10 3166 \#1 & M5.0V    & 14.71 & 13.36 & 11.78 & 14.43 & 13.03 & 11.22 & 1 &  9.51 &  8.97 &  8.64 &  16.4 &  10.1 & 11 &  12.5 &   2.0 & 12 \\
HD 033636 \#1    & M1.0V    & 20.56 & 18.17 &       & 19.31 & 18.43 & 17.37 & 1 & 16.26 & 15.63 & 15.16 & 608.5 & 162.9 &  6 & 738.9 & 162.3 & 12 \\
HD 041004 \#1    & M0.5V-VI & 18.90 & 16.87 & 15.76 & 17.89 & 16.91 & 16.05 & 1 & 15.06 & 14.50 & 14.16 & 414.0 & 119.1 & 11 & 557.4 & 103.3 &  9 \\
HD 072659 \#1    & M3.0V    & 20.21 & 18.05 & 16.43 & 18.91 & 18.02 & 16.53 & 1 & 15.31 & 14.67 & 14.30 & 293.0 &  82.5 & 11 & 368.6 &  99.2 & 12 \\
HD 114783 \#1    & Early K  & 10.60 &  9.32 &  8.92 &  9.78 &  9.31 &  8.90 & 2 &  8.32 &  7.90 &  7.79 &  20.2 &   5.4 &  3 &  54.0 &   9.3 &  2 \\
\enddata
\end{deluxetable}
\clearpage
\setlength{\voffset}{0mm}
\figcaption[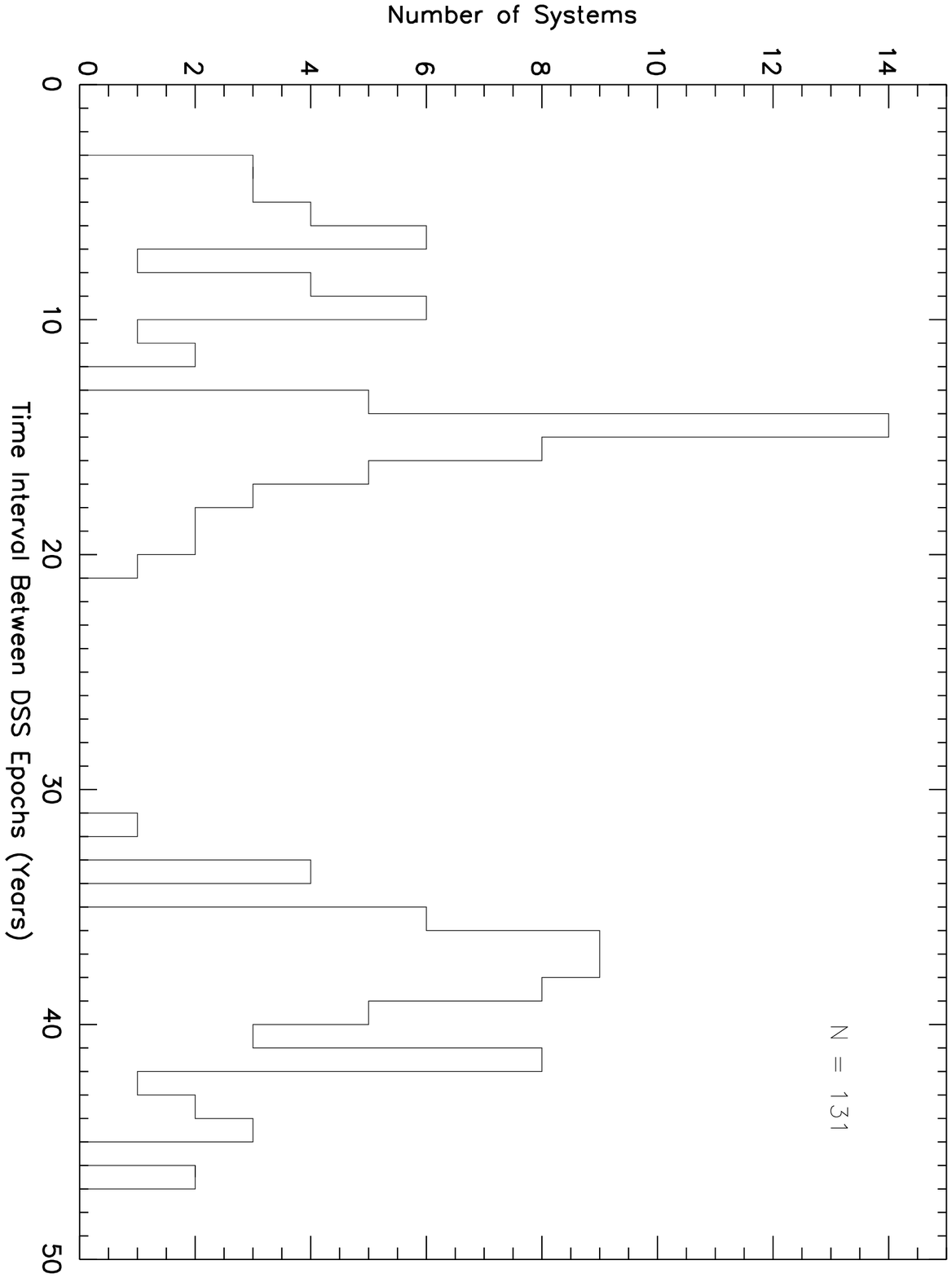]{Histogram of time intervals between DSS epochs for
the exoplanet sample.
\label{DSS-hist}}

\figcaption[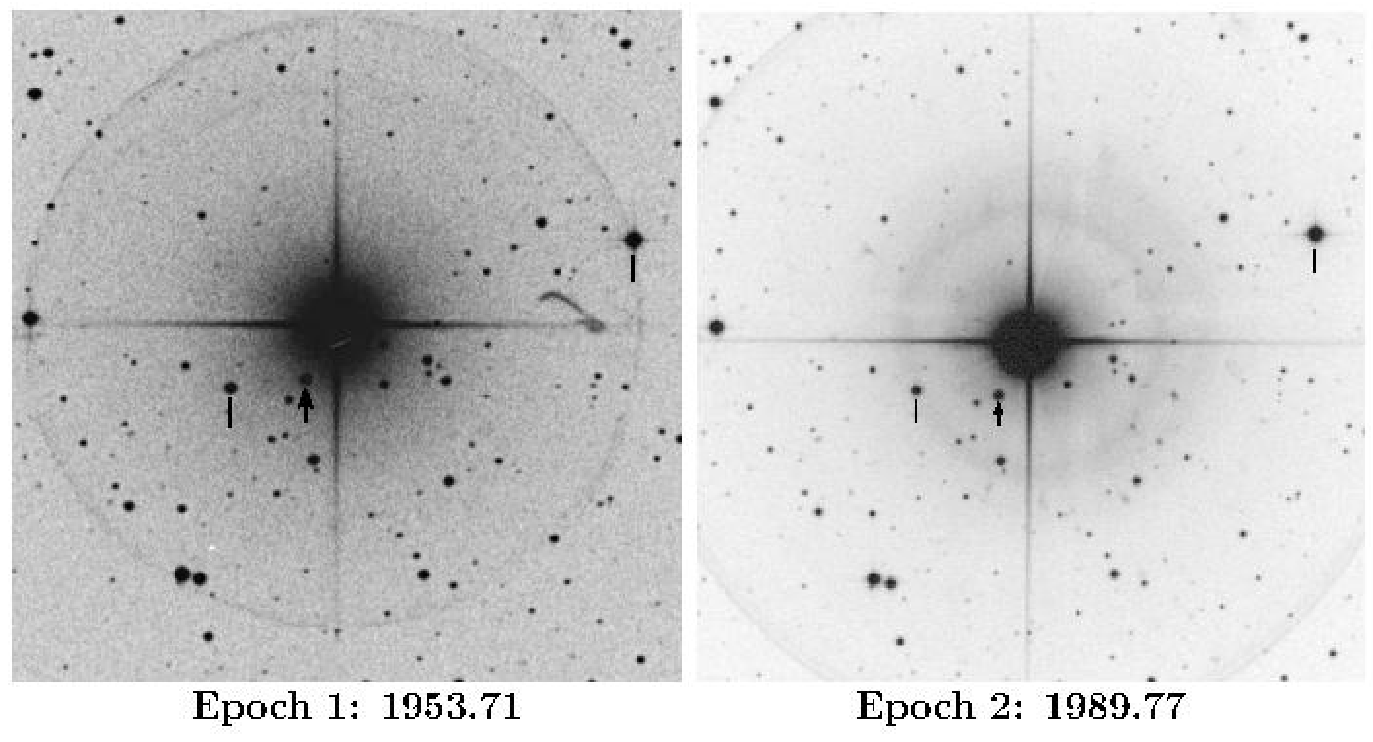]{DSS images from two epochs for HD 9826.  The
  10$\arcmin$ square images have north up and east to the left.  WDS
  lists components B \& C (marked by lines), which are background
  stars.  WDS component D (marked by an arrow), however, is a CPM
  companion.  The primary's $\mu$ = 0\farcs42 yr$^{-1}$ at 204\sdeg.
\label{WDS-CPM}}

\figcaption[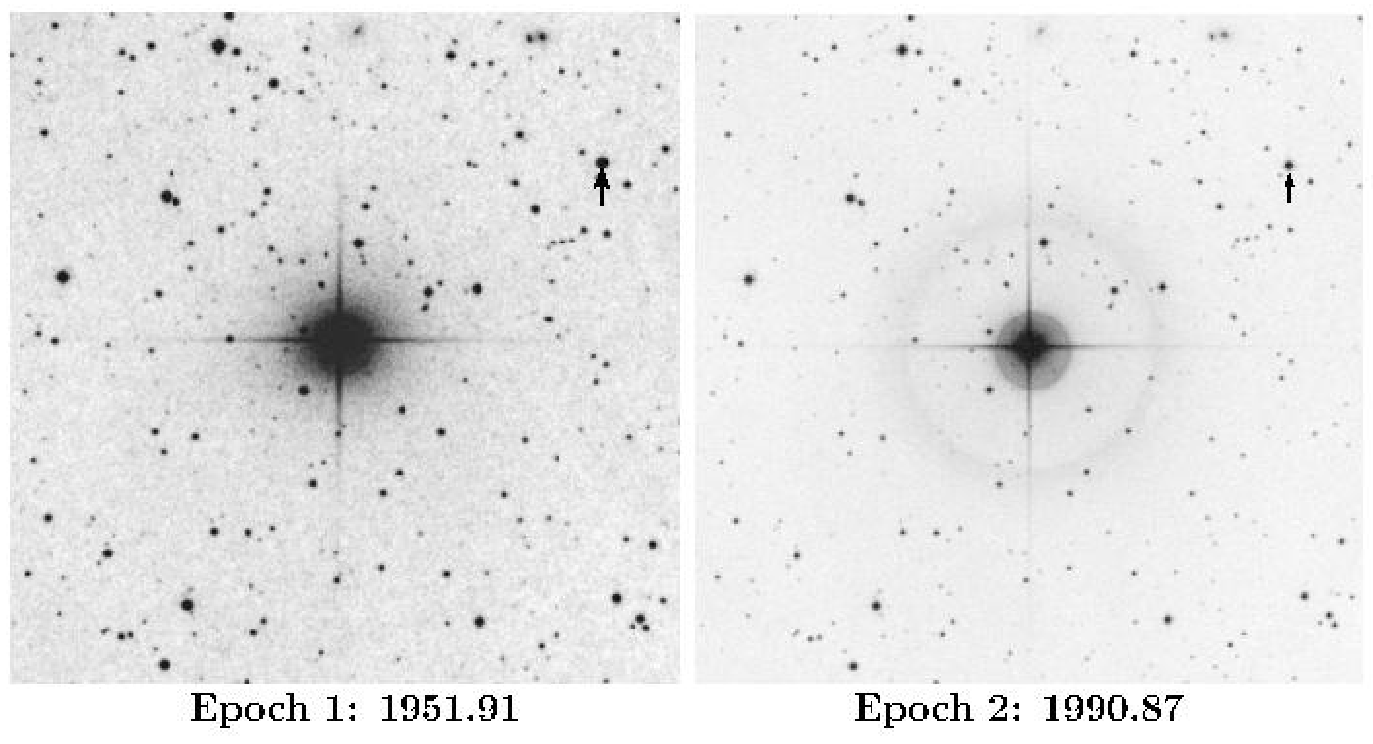]{New stellar companion to exoplanet host HD 38529.
The 10$\arcmin$ square DSS images have north up and east to the
left. The companion, marked by arrows, is at an angular separation of
284$\arcsec$ at 305\sdeg~from the primary, which is at the center of
the images.
\label{new38529}}

\figcaption[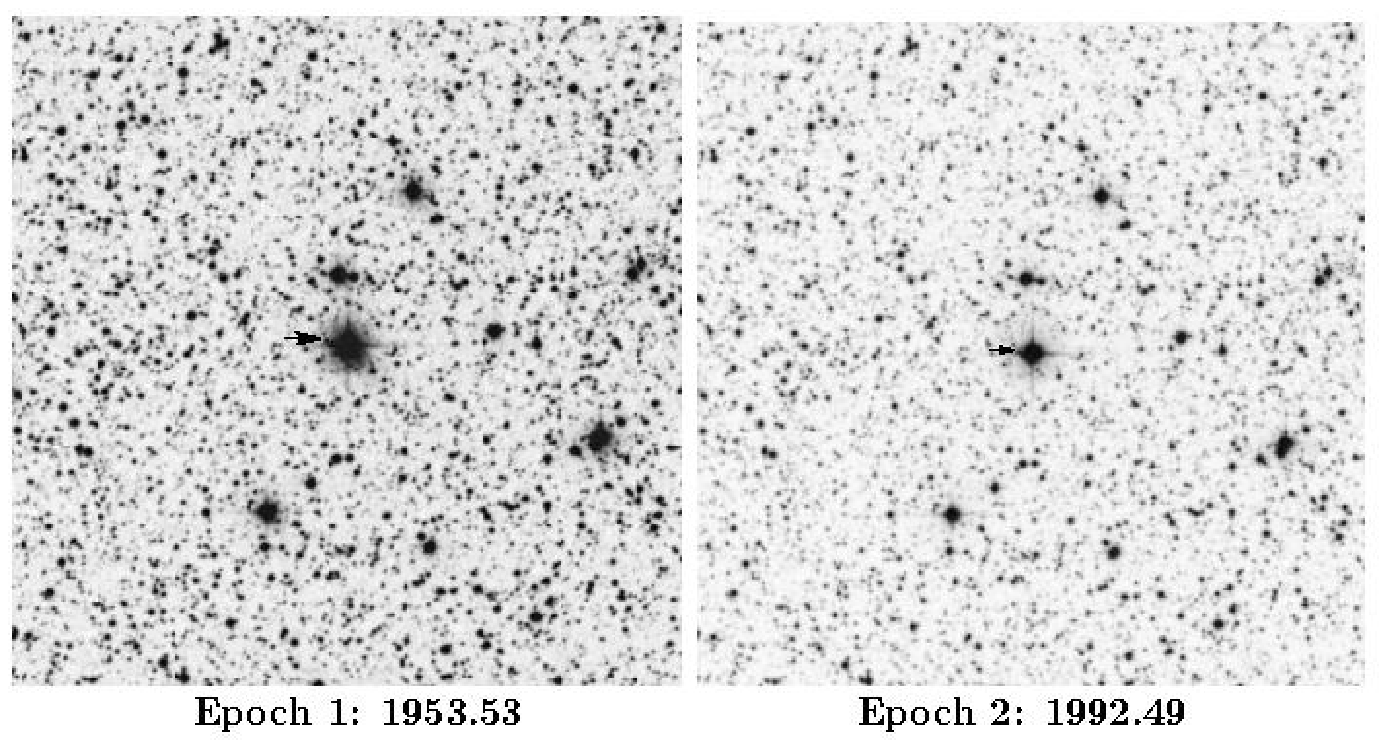]{New stellar companion to exoplanet host HD 188015.
The 10$\arcmin$ square DSS images have north up and east to the
left. The companion, marked by arrows, is at an angular separation of
13$\arcsec$ at 85\sdeg~from the primary, which is the bright source at 
the center of the images.
\label{new188015}}

\figcaption[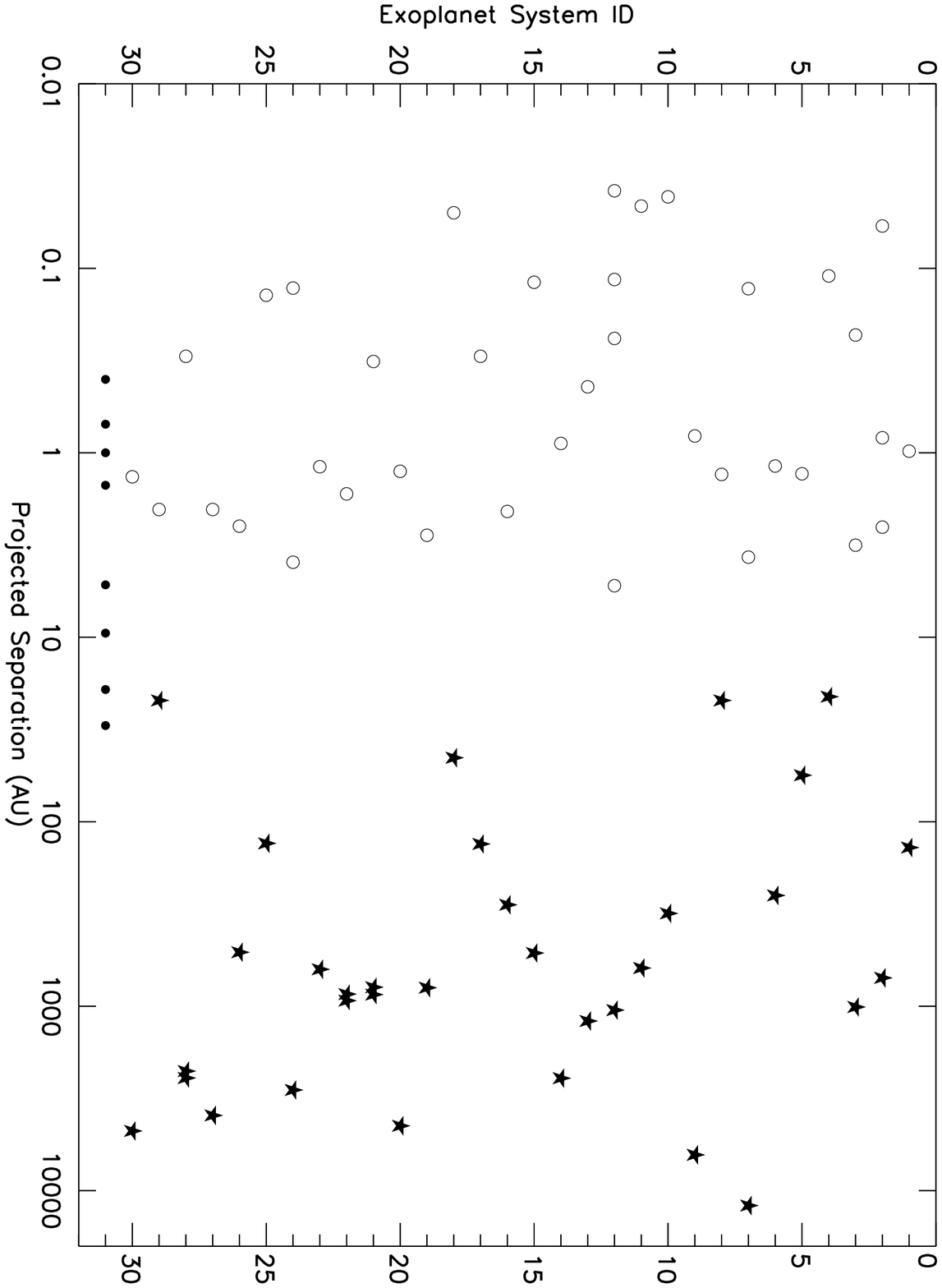]{Orbits of planets and stars in exoplanet systems
with stellar companions.  The exoplanet host stars are at a position
of zero AU.  Open circles represent planets and stars represent stars.
Points will tend to move right because of orbital inclination and
projection effects.  Separations between the components of the three
binary companions is exaggerated to be able to distinguish the binary
components on the plot.  For comparison, the positions of the eight
planets of our Solar System are shown at the bottom as filled circles.
\label{Orbits}}

\figcaption[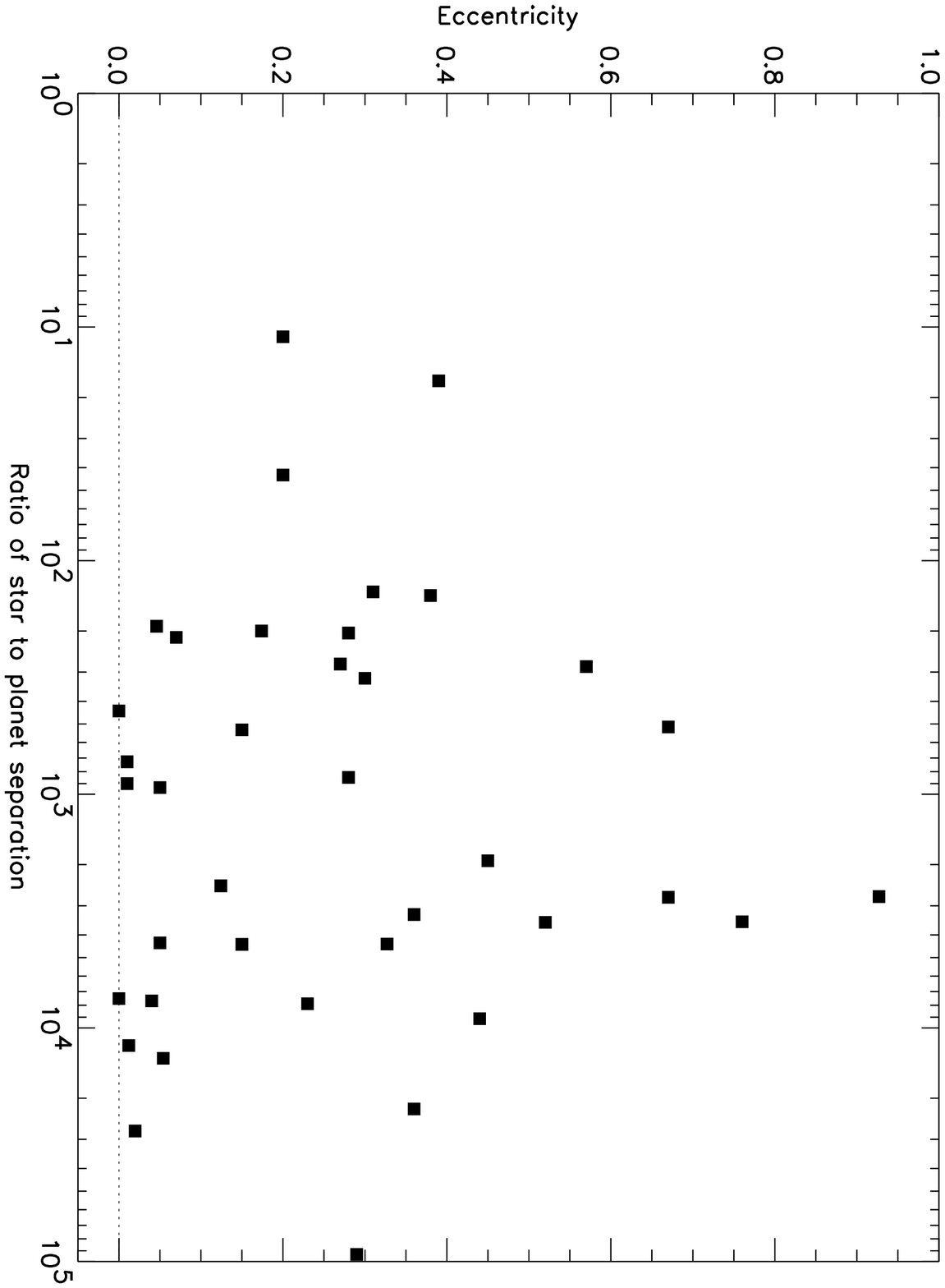]{Eccentricity of planetary orbits as a function of
proximity of the stellar companion.  The ratio is computed using
projected stellar separation and a $\sin{i}$ of the planetary orbit.
\label{eccrat}}

\figcaption[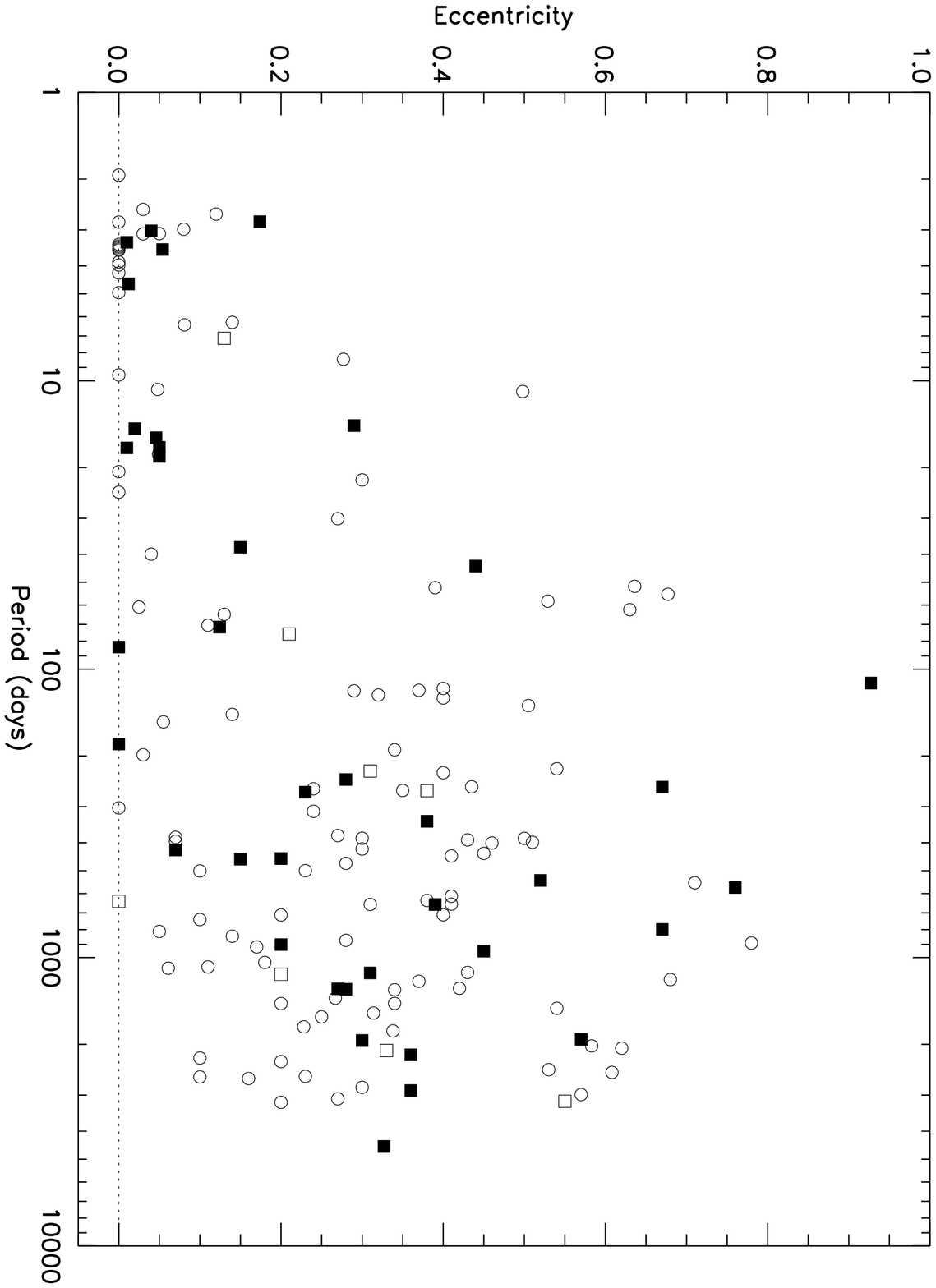]{Period-eccentricity diagram for planets orbiting
single stars (open circles) and planets in systems with more than one
star (open squares for candidate multiplicity, and filled squares for 
confirmed multiplicity).
\label{eccper}}


\begin{figure}
\epsscale{0.6}\plotone{f1.eps}
\centerline{f1}
\end{figure}
\clearpage

\begin{figure}
\includegraphics[angle=90,width=30cm]{f2.eps}
\centerline{f2}
\end{figure}
\clearpage

\begin{figure}
\includegraphics[angle=90,width=30cm]{f3.eps}
\centerline{f3}
\end{figure}
\clearpage

\begin{figure}
\includegraphics[angle=90,width=30cm]{f4.eps}
\centerline{f4}
\end{figure}
\clearpage

\begin{figure}
\epsscale{0.8}\plotone{f5.eps}
\centerline{f5}
\end{figure}
\clearpage

\begin{figure}
\epsscale{1.0}\plotone{f6.eps}
\centerline{f6}
\end{figure}
\clearpage

\begin{figure}
\epsscale{1.0}\plotone{f7.eps}
\centerline{f7}
\end{figure}

\end{document}